\documentclass[a4paper,twocolumn,superscriptaddress,11pt,accepted=2022-06-07]{quantumarticle}

\pdfoutput=1

\usepackage[numbers,sort&compress]{natbib}
\usepackage[normalem]{ulem}
\usepackage{natbib}
\usepackage[T1]{fontenc}
\usepackage{float}
\usepackage{rotating}
\usepackage{graphicx}
\usepackage{xcolor}

\usepackage{latexsym}
\usepackage{amsmath}
\usepackage{amssymb}
\usepackage{amsfonts}
\usepackage{amsthm}
\usepackage{mathrsfs}
\usepackage{mathtools}
\usepackage{verbatim}
\usepackage{psfrag}
\DeclareMathAlphabet{\mathrsfs}{U}{rsfs}{m}{n}
\DeclareMathAlphabet{\mathpzc}{OT1}{pzc}{m}{it}
\DeclareMathAlphabet{\matheus}{U}{eus}{m}{n}
\DeclareMathAlphabet{\mathbbold}{U}{bbold}{m}{n}

\def\one{\leavevmode\hbox{\small1\normalsize\kern-.33em1}}

\newcommand{\ba}{\begin{eqnarray}}
\newcommand{\ea}{\end{eqnarray}}
\newcommand{\ban}{\begin{eqnarray*}}
\newcommand{\ean}{\end{eqnarray*}}

\begin{document}

\title{Platonic Bell inequalities for all dimensions}

\author{K\'aroly F. P\'al}
\email{kfpal@atomki.hu}
\affiliation{Institute for Nuclear Research, P.~O. Box 51, H-4001 Debrecen, Hungary}

\author{Tam\'as V\'ertesi}
\email{tvertesi@atomki.hu}
\affiliation{MTA Atomki Lend\"ulet Quantum Correlations Research Group, Institute for Nuclear Research, P.~O. Box 51, H-4001 Debrecen, Hungary}	
	
\begin{abstract}
In this paper we study the Platonic Bell inequalities for all possible dimensions. There are five Platonic solids in three dimensions, but there are also solids with Platonic properties (also known as regular polyhedra) in four and higher dimensions. The concept of Platonic Bell inequalities in the three-dimensional Euclidean space was introduced by Tavakoli and Gisin [Quantum 4, 293 (2020)]. For any three-dimensional Platonic solid, an arrangement of projective measurements is associated where the measurement directions point toward the vertices of the solids. For the higher dimensional regular polyhedra, we use the correspondence of the vertices to the measurements in the abstract Tsirelson space [B. Tsirelson, J. Soviet Math. 36, 557 (1987)]. We give a remarkably simple formula for the quantum violation of all the Platonic Bell inequalities, which we prove to attain the maximum possible quantum violation of the Bell inequalities, i.e. the Tsirelson bound.
To construct Bell inequalities with a large number of settings, it is crucial to compute the local bound efficiently. In general, the computation time required to compute the local bound grows exponentially with the number of measurement settings. We find a method to compute the local bound exactly for any bipartite two-outcome Bell inequality, where the dependence becomes polynomial whose degree is the rank of the Bell matrix. To show that this algorithm can be used in practice, we compute the local bound of a 300-setting Platonic Bell inequality based on the halved dodecaplex. 
In addition, we use a diagonal modification of the original Platonic Bell matrix to increase the ratio of quantum to local bound. In this way, we obtain a four-dimensional 60-setting Platonic Bell inequality based on the halved tetraplex for which the quantum violation exceeds the maximum quantum violation of the famous Clauser-Horne-Shimony-Holt Bell inequality. This is the first example of a Platonic Bell inequality exceeding the $\sqrt 2$ ratio. 		
\end{abstract}
	

\maketitle
	
	
\section{Platonic solids in quantum physics}\label{PlatQI}
Platonic solids~\cite{coxeter} -- convex polyhedra with equivalent faces composed of congruent convex regular polygons -- have been recently studied in the context of Bell nonlocality~\cite{Bell64,bellreview}. Specifically, Tavakoli and Gisin~\cite{TavGis} constructed bipartite two-outcome Bell inequalities~\cite{cirel80,Tsirelson87} whose maximal violation is obtained with projective measurements pointing toward the vertices of three-dimensional Platonic solids. These inequalities are called Platonic Bell inequalites~\cite{TavGis}. 

In addition to Platonic solids, semi-regular polyhedra, namely Archimedean solids~\cite{coxeter}, were studied in Ref.~\cite{TavGis}. A recent development is the realization that starting from the fact that Platonic and Archimedean solids may be generated as orbits of representations of groups makes it possible to derive very interesting general properties of the Bell inequalities constructed~\cite{bolonek}. The group-theoretical approach also allows one to derive further inequalities sharing these properties based on three-dimensional bodies even without apparent symmetries.

All the constructed Bell expressions involve only joint correlation terms and they are denoted correlation-type Bell expressions~\cite{Tsirelson87} (also known as XOR games~\cite{Cleve}). The maximum quantum value of these Bell expressions can be achieved by maximally entangled states~\cite{Tsirelson87}. The ratio $Q/L$ between the optimal quantum value $Q$ for such a Bell expression and its local bound $L$ quantifies the amount of isotropic noise $p=1-(L/Q)$ which must be added to the maximally entangled state such that the quantum value becomes smaller than $L$. Therefore the $Q/L$ ratio is relevant from an experimental point of view as well. As a matter of fact, no Bell inequalities arising from these three-dimensional solids were found~\cite{TavGis,bolonek} which have a higher $Q/L$ ratio than $\sqrt 2$, i.e., exhibit higher noise resistance to the isotropic noise than the famous Clauser-Horne-Shimony-Holt (CHSH) Bell inequality~\cite{CHSH}. Note, however, that there exist correlation-type Bell inequalities overcoming the ratio $Q/L=\sqrt 2$ using maximally entangled states which are not based on Platonic Bell inequalities, see e.g.~\cite{fishburn,acin06}. The first example using a two-qubit maximally entangled state appeared in 2008 in Ref.~\cite{vertesi}.

Our present study is built upon and motivated by the recent work of Tavakoli and Gisin~\cite{TavGis}. We generalize the notion of Platonic Bell inequalities to $d\neq 3$ dimensional Euclidean spaces. Obviously, if $d>3$, this space can not be identified with the physical space such that the measurements point towards the vertices of the solids. Nevertheless, even in such cases the generalization does correspond to genuine quantum Bell scenarios. We have studied all pairs of Platonic solids in all dimensions. In contrast to previous works, we have also considered different local orientations of these solids, especially to determine which relative orientation gives the maximum $Q/L$ ratio. We have proven that the quantum value is independent of the relative orientation and can be given in all cases by a very simple formula. We have found a method to efficiently calculate the local bound when the rank of the Bell matrix (which in this case is just $d$) is much smaller than the number of measurement settings. Unfortunaly, we still could not find a single Bell inequality where $Q/L$ is larger than $\sqrt 2$. Next, we have also studied Platonic Bell inequalities where the main diagonal entries of the Bell matrix have been modified by a constant term. We have shown that this modification does not change the measurement configuration corresponding to the maximum quantum violation. Therefore, the diagonally modified inequalities are still considered as Platonic Bell inequalities. Finally, starting from the 120-vertex tetraplex, with this diagonal modification we have found Platonic Bell inequalities with $Q/L>\sqrt 2$. Hence, these Bell inequalities are more tolerant to isotropic noise than the CHSH inequality. In fact, they require a state space of dimension $4\times 4$ to reach the maximum quantum violation, i.e., the Tsirelson bound~\cite{cirel80}. 

Besides Bell nonlocality, there is a number of other uses of Platonic solids in quantum physics. Let us give just a few other examples. In Refs.~\cite{bennet2012,saunders}, the directions of quantum measurements are chosen along the vertices of Platonic solids to demonstrate Einstein-Podolsky-Rosen steering. In Ref.~\cite{decker}, quantum circuits have been devised where the measurement directions point to vertices of the Platonic solids. In Ref.~\cite{jeong}, different types of single qubit private quantum channels have been constructed and linked to three-dimensional Platonic solids. In a more recent work~\cite{lee} four-dimensional Platonic solids were connected to qutrit-based private quantum channels. Such regular polyhedra were also used in reference frame alignment~\cite{kolen} and quantum hashing protocols~\cite{burr}. More recently Platonic solids were considered in quantum entanglement theory~\cite{latorre} and quantum contextuality~\cite{xiao}.

The structure of the paper is as follows. Section~\ref{introplaton} gives an introduction to correlation-type Bell inequalities, and more specifically to Platonic Bell inequalities introduced in Ref.~\cite{TavGis}. Section~\ref{Platsold} summarizes the basic properties of Platonic solids in all the possible dimensions. In Section \ref{Qvalspec} the maximum quantum value, that is the quantum bound of a special class of Bell inequalities is given. This class is proved to include all Platonic (and Archimedean) inequalities. We derive a remarkably simple formula for the quantum violation of all members of this class. We note that the same formula~(\ref{eq:Th1}) for the quantum bound in the framework of group theory has been obtained independently in Ref.~\cite{bolonek}. However, more conditions have been used than necessary for the derivation, and no proof has been provided that the formula also defines the globally optimal quantum value, i.~e., that it reaches the Tsirelson bound of the Bell inequality. In contrast, we show that the choice of the measurement vectors is optimal, proving that the measurement configuration allows us to reach the Tsirelson bound.

To construct Bell inequalities with a large number of settings, it is crucial to compute the local bound efficiently. In general, the computation time required to compute the local bound grows exponentially with the number of measurement settings. In Section~\ref{Lbound}, we give a numerical method that allows the computation of the local bound in polynomial time (instead of exponential scaling) in the number of measurement settings for any bipartite two-outcome Bell inequality with constant rank of the Bell matrix. The dependence is polynomial in terms of the rank of the Bell matrix, in particular, the degree of the polynomial equals the rank of the Bell matrix. The considered bipartite Bell inequalities include all correlation-type inequalities and all Platonic Bell inequalities as a subset of correlation-type inequalities. In the next sections we apply the developed tools to Platonic Bell inequalities in all possible dimensions: In Section~\ref{Twodim}, we consider Platonic Bell inequalities based on regular polygons (that is, the case of $d=2$). The three and higher dimensional cases are discussed in Sections~\ref{Locstratdge3} and \ref{MinLocsdge3}. In particular, the local strategies for all $d\geq 3$ are given in Section~\ref{Locstratdge3} and the computation of the local bounds for $d\geq 3$ is carried out in Section~\ref{MinLocsdge3}. Then, in Section~\ref{belldiag}, we discuss Platonic Bell inequalities where the main diagonal entries of the Bell matrix are modified with a constant term. In this way, we find (diagonally modified) Platonic Bell inequalities that are more robust to isotropic noise than the celebrated CHSH-Bell inequality. The obtained results are summarized in Section~\ref{disc}. In Appendix~\ref{EuHil} we discuss the known relationship between the Euclidean and Hilbert space descriptions whose existence shows that higher dimensional generalizations do indeed correspond to quantum Bell scenarios. The proof of special properties of our classes of Bell inequalities is given in Appendix~\ref{specpropproof}.	
	
\section{Introduction to Platonic Bell inequalities}\label{introplaton}
Tavakoli and Gisin \cite{TavGis} have constructed two-party correlation-type Bell inequalities~\cite{Tsirelson87} whose maximal quantum value can be reached by measurements performed on a pair of one-half spins along directions of the vertices of Platonic solids. In a two-party Bell scenario parts of identically prepared physical systems are distributed between two distant parties, say Alice and Bob. Each of them independently chooses a measurement setting from a pre-decided set and performs the corresponding measurement on her/his subsystem. A Bell inequality expresses the fact that, given the measurement settings chosen by the parties, a linear combination of the conditional probabilities of the possible outcomes cannot be greater than a certain value, the so-called local bound, in any locally realistic theory. Quantum theory allows for the violation of many Bell inequalities~\cite{bellreview}.

Correlation-type Bell inequalities~\cite{Tsirelson87} involve binary, that is two-outcome measurements. Let the two outcomes be $+1$ and $-1$. In the present case, when the system is a pair of one-half spins, the spin projections are measured along pre-determined spatial directions. The possible result of such a measurement is either $+\hbar/2$ (outcome $+1$) or $-\hbar/2$ (outcome $-1$). The general form of such an inequality may be written as:
\begin{equation}
\sum_{i=1}^{m_A}\sum_{j=1}^{m_B}M_{ij}\langle{a_ib_j}\rangle\leq L,
\label{eq:corbell}
\end{equation}
where $m_A$ and $m_B$ are the number of measurement settings, while $a_i=\pm 1$ and $b_j=\pm 1$ are the outcome of measurement $i$ and $j$ of Alice and Bob, respectively. The $M$ is the Bell matrix whose real elements are the Bell coefficients. The expectation value $\langle{a_ib_j}\rangle$ is $p(11|ij)+p(-1-1|ij)-p(1-1|ij)-p(-11|ij)$, that is it can be expressed as a linear combination of $p(a_ib_i|ij)$ conditional probabilities. The value $L$ is the maximum value the expression may take considering all possibilities with deterministic outcomes (when the outcomes are the same in all rounds with the same choices of settings), which is the maximum value allowed by any locally realistic theory.

In quantum mechanics $\langle{a_ib_j}\rangle=\mathrm{Tr}(\rho\hat a_i\otimes\hat b_j)$ in case of projective measurements, where $\hat a_i$ and $\hat b_j$ are the Hermitian measurement operators with eigenvalues $\pm 1$ corresponding to the $i$th and $j$th measurement of Alice and Bob, respectively, while $\rho$ is the density operator of the system. For a pure state $|\psi\rangle$ this becomes $\langle\psi|\hat a_i\otimes\hat b_j|\psi\rangle$. The maximal quantum value, that is, the quantum bound, can always be achieved with a pure state, in the case of a correlation-type Bell inequality with a maximally entangled one~\cite{Tsirelson87,acin06}. The quantum bound can be determined by semidefinite programming applying the Navascués-Pironio-Acín (NPA) hierarchy of Ref.~\cite{npa} at its first level. 

For a qubit the most general non-degenerate measurement operator is $\alpha_x\hat\sigma_x+\alpha_y\hat\sigma_y+\alpha_z\hat\sigma_z\equiv\vec\alpha\cdot\vec\sigma$, where $\hat\sigma_x$, $\hat\sigma_y$ and $\hat\sigma_z$ are the Pauli-operators. If the qubit corresponds to a spin, then the unit vector $\vec\alpha$ points towards the direction in the physical space along which the spin projection is measured. It can be shown (for example \cite{TavGis}) that if the measurement by Alice and Bob is performed on a pair of Bell state $|\Phi^+\rangle =(|00\rangle+|11\rangle)/\sqrt{2}$ along $\vec\alpha$ and $\vec\beta$, respectively, then the expectation value of the product of the measurement outcomes $\langle\Phi^+|(\vec\alpha\vec\sigma)\otimes(\vec\beta\vec\sigma)|\Phi^+\rangle=\vec\alpha\cdot\vec\beta^*$, where $\beta^*_x\equiv\beta_x$, $\beta^*_y\equiv-\beta_y$ and $\beta^*_z\equiv\beta_z$. Therefore, the quantum value of the Bell inequality with the $|\Phi^+\rangle$ state is:
\begin{equation}
Q=\sum_{i=1}^{m_A}\sum_{j=1}^{m_B}M_{ij}\vec a_i\cdot\vec b_j,
\label{eq:vecbell}
\end{equation}
where $\vec a_i$ is Alice's $i$th measurement direction, while $\vec b_j$ is Bob's $j$th measurement direction reflected through the $x-z$ plane.

The systematic method to construct the desired Bell inequalities Tavakoli and Gisin \cite{TavGis} proposed was to take $M_{ij}=\vec A_i\cdot\vec B_j$, where $\vec A_i$ and $\vec B^*_j$ are the unit vectors pointing towards the desired measurement directions in Alice's, and  Bob's side, respectively, namely towards vertices of Platonic bodies. Then the quantum value for the resulting Bell inequality with these measurement directions is $\sum_{ij}(\vec A_i\cdot\vec B_j)^2$. Although all terms in the expression are positive, it itself does not ensure that the desired measurement directions are really optimal. For a general choice of vectors $\vec A_i$ and $\vec B_j$ this is usually not true. However, the authors checked all pairs of Platonic solids, although only with one relative orientation for each pair, by calculating the quantum bound using the NPA method~\cite{npa}, and found that the desired measurement directions were optimal for all of cases they checked. We will show here that all Platonic solids have certain property that ensures this, moreover, the form of the quantum bound is very simple and independent of the relative orientation of the two bodies. However, the local bound, and hence the maximum violation of the inequality does depend on this. We will discuss this dependence in the present paper.

Tavakoli and Gisin \cite{TavGis} have considered only the classical, three-dimensional Platonic solids. The objects corresponding to the Platonic solids in two dimensions are the regular convex polygons. There is an infinite number of them. If the system is a pair of one-half spins, the Bell inequality, which is constructed according to Ref.~\cite{TavGis} from two such polygons in the same plane, will be maximally violated by measurements towards the vertices of the polygons. We will explore this generalization in Section~\ref{Twodim}. There are generalizations of the Platonic solids in higher dimensions as well. In four dimensions there are six such objects, while in each higher dimensional space there are only three of them. Due to a theorem by Tsirelson \cite{Tsirelson87}, for any Euclidean space there exists a Hilbert space $H$ and a state $|\Phi\rangle$ in $H\otimes H$ such that for all unit vectors $\vec a$ and $\vec b$ in the Euclidean space there exist measurement operators $\hat a$ and $\hat b$ in $H$ such that $\vec a\cdot\vec b=\langle\Phi|\hat a\otimes\hat b|\Phi\rangle$. For a three dimensional Euclidean space the Hilbert space is the qubit space and the operators are $\vec a\cdot\vec\sigma$ and $\vec b^*\cdot\vec\sigma$. In Appendix~\ref{EuHil} we discuss the relationship between the Euclidean and Hilbert space description. The considerations there are based on the explicit construction for all dimensions given for example in Refs.~\cite{acin06,vp08}. From the existence of such a construction, it follows that for any set of $\vec a_i$ and $\vec b_j$ vectors in any Euclidean space, the expression $\sum_{ij}M_{ij}\vec a_i\cdot\vec b_j$ will give the quantum value of the Bell expression with certain measurements performed on certain bipartite quantum system. This means that the generalization of the construction of the Bell inequalities to higher dimensional spaces does make sense, the result will correspond to a genuine Bell scenario. We will call the $\vec a_i$ and $\vec b_j$ the measurement vectors. 

Unfortunately, the beauty of those measurement settings corresponding to highly symmetric arrangements in the physical space is necessarily lost in more than three dimensions. At the same time, in experimental realizations of Bell scenarios the quantum systems are typically not pairs of one-half spins even if they are pairs of qubits. In that case symmetry would only show up in an abstract space anyway, just like in the generalization to higher dimensions.

In more than two dimensions some Platonic solids have too many vertices, such that the calculation of the local bound of Bell inequalities constructed from them in the straightforward way may not be feasible. We have found a method which makes it possible to calculate the local bound exactly whenever the Bell coefficients are given as scalar products of low dimensional Euclidean vectors, even if the number of measurement settings is large. As the vectors need not be unit vectors in this case, the method may be useful whenever the rank of the Bell matrix of a correlation-type inequality is low. In addition, we show that the computation of the local bound can be done efficiently for any bipartite two-outcome Bell inequality with low Bell matrix rank.

\section{Platonic solids in $d$ dimensions}\label{Platsold}
In this section we briefly summarize some well-known facts about Platonic bodies~\cite{coxeter}. All this knowledge can even be found on Wikipedia.

For $d>4$ there are just three types of Platonic solids: the simplex, the cross polytope and the hypercube. A simplex has $d+1$ vertices, and if the vectors pointing towards those vertices are normalized, the scalar product of each pair of them is $-1/d$. A cross polytope has $2d$ vertices. In its standard orientation, which is the most convenient one, these vertices are given by $\vec e_i$ and $-\vec e_i$, where $\vec e_i$ are the coordinate vectors. The $d$-dimensional hypercube or simply $d$-cube has $m=2^d$ vertices $\vec V_i$, whose each coordinate $V_{ij}$ in the standard orientation is either $+1/\sqrt{d}$ or $-1/\sqrt{d}$, such that all variations of the signs are covered. Cross polytopes and hypercubes are centrally symmetric, while simplices are not.

In two dimensions there are an infinite number of Platonic bodies, they are the regular convex polygons. An $n$-sided regular polygon is centrally symmetric if and only if $n$ is even. From our point of view if $n$ is odd, the $n$-sided and the $2n$-sided polygons behave very similarly. The two-dimensional simplex is the regular triangle, while both the cross polytope and the 2-cube is the square, their standard orientation differs by an angle of $\pi/4$.

In three dimensions the simplex is the tetrahedron, the cross polytope is the octahedron, while the 3-cube is simply the cube. If we reflect the vertices of the tetrahedron to its center and add the reflected points as new vertices, we get a cube. As we will see later, for this reason the two bodies will behave equivalently from our point of view. There are two additional three-dimensional Platonic bodies, the icosahedron and the dodecahedron having $12$ and $20$ vertices, respectively. Both are centrally symmetric.

In four dimensions the simplex is the pentachoron, but it is also referred to as 5-cell, pentatope, pentahedroid, or tetrahedral pyramid. The cross polytope is the hexdecahedroid, also called 16-cell or hexadecachoron, while the 4-cube is the tesseract, whose alternative names are 8-cell, octachoron, octahedroid, cubic prism, and tetracube. There are three additional Platonic bodies in $d$=4: the octaplex (24-cell, icositetrachoron, icosatetrahedroid, octacube, hyper-diamond, polyoctahedron), the tetraplex (600-cell, hexacosichoron, hexacosihedroid, polytetrahedron) and the dodecaplex (120-cell hyperdodecahedron, polydodecahedron, hecatonicosachoron, dodecacontachoron, hecatonicosahedroid) with 24, 120 and 600 vertices, respectively. These three bodies are centrally symmetric. We note that in the name $n$-cell for each $4$-dimensional body $n$ refers to the number of facets, while for us the number of vertices is relevant.

\section{Quantum bound for the special cases}\label{Qvalspec}
{\bf Observation:} The set of unit vectors $\vec V_i$ pointing towards the vertices of a Platonic solid has the property that the columns of the matrix whose elements $V_{ij}\equiv(\vec V_i)_j$ are the components of the vectors are orthogonal to each other and have equal norm, that is $\sum_{i=1}^m V_{ij}V_{ik}=\delta_{jk}m/d$, where  $m$ is the number of vertices, $d$ is the dimensionality of the space and $j,k=1,\dots d$. In other words, matrix $V$ is proportional to a semiorthogonal matrix.

The value of the norm of the columns follows from the unit lengths of $\vec V_i$. If $\vec V_i$ has the property, then
\begin{equation}
\sum_{i=1}^m (\vec V_i\cdot \vec x)\vec V_i=\sum_{i=1}^m\sum_{j,k=1}^d(V_{ij}x_j)(V_{ik}\vec e_k)=\frac{m}{d}\vec x,
\label{eq:vproperty1}
\end{equation}
where $\vec x$ is any vector in the $d$-dimensional space, and $\vec e_k$ are the basis vectors. From this it follows that if $\vec y$ is also a $d$-dimensional vector, then
\begin{equation}
\sum_{i=1}^m (\vec V_i\cdot \vec x)(\vec V_i\cdot \vec y)=\frac{m}{d}\vec x\cdot \vec y.
\label{eq:vproperty2}
\end{equation}
From the equation above it is clear that the property is independent of the choice of the basis vectors ($\vec x$ and $\vec y$ may be members of another basis), therefore, the property is invariant to any orthogonal transformation.

Now let the Bell coefficients be $M_{ij}=\vec A_i\cdot \vec B_j$ ($i=1,\dots,m_A$; $j=1,\dots,m_B$). We will refer to $\vec A_i$ and $\vec B_j$ as the construction vectors. Furthermore, let either $\vec A_i$ or $\vec B_j$ obey the constraint given in the Observation. Then the quantum value one gets when the measurement vectors are the same as the respective construction vectors, i.e.\ $\vec a_i=\vec A_i$ and $\vec b_i=\vec B_i$ is
\begin{equation}
\sum_{i=1}^{m_A}\sum_{j=1}^{m_B}(\vec A_i\cdot \vec B_j)(\vec A_i\cdot \vec B_j)=\frac{m_Am_B}{d},
\label{eq:Th1}
\end{equation}
which readily follows from Eq.~(\ref{eq:vproperty2}) and from the fact that $\vec A_i$ and $\vec B_j$ are unit vectors. It is trivial that the quantum value calculated this way does not depend on the relative orientation of the two sets of vectors, it is invariant to orthogonal transformations applied on either side. The same formula has been obtained for three dimensions independently in Ref.~\cite{bolonek}. To derive it, the authors assumed that the vectors on both sides correspond to orbits of the same group generated by the same three-dimensional irreducible representation. Eqs.~(15) and (16) of the paper show that if the vectors generated this way, they will obey the constraint given in the Observation on both sides, which is unnecessary. As we have shown, if the condition holds on either side, then Eq.~\ref{eq:Th1} is true. However, this quantum value will usually be below the quantum bound unless the constraint is obeyed on both sides, in which case the quantum bound is reached, as we will prove later. If the vectors are generated as orbits, as in Ref.~\cite{bolonek}, it is not necessary to use the same representation of the same group on both sides; the quantum bound will be given by the simple formula even if different groups are chosen.

The quantum values given by Tavakoli and Gisin \cite{TavGis} for 3-dimensional Platonic solids all agree with $m_Am_B/3$ given above. They have shown for each pairs of solids that taking the measurement vectors equal to the construction vectors is actually an optimum choice, which gives the maximum quantum value. For that they used the NPA hierarchy \cite{npa}. Proof was necessary, because although such a choice must clearly be a good one, as each term in the Bell-expression gives a positive contribution, but actually this is usually not the optimum one if the construction vectors are chosen differently.

Now we will show that if both $\vec A_i$ and $\vec B_i$ obey the constraint, this choice of the measurement vectors is optimal by proving that the quantum value agrees with the upper bound derived by Epping {\it et al.}\ in Ref.~\cite{EppKamBru1} for two-party binary-outcome correlation-type Bell inequalities. The bound is given as $\sqrt{m_Am_B}||g||_2$, where $||g||_2$ is the maximal singular value of the Bell matrix. The Bell matrix can be rewritten as:
\begin{equation}
M_{ij}=\vec A_i\cdot \vec B_j=\sum_{k=1}^dA_{ik}B^T_{kj}=\sum_{k=1}^d\bar A_{ik}S_{kk}\bar B^T_{kj},
\label{eq:Bellsing}
\end{equation}
where $A_{ik}\equiv(\vec A_i)_k$, $B_{jk}\equiv(\vec B_j)_k$, $\bar A_{ik}\equiv\sqrt{d/m_A}A_{ik}$, $\bar B_{jk}\equiv\sqrt{d/m_B}B_{jk}$ and $S$ is a diagonal matrix with nonzero elements $S_{kk}=\sqrt{m_Am_B}/d$. The constraint for $\vec A_i$ and $\vec B_i$ means that matrices $\bar A$ and $\bar B$ are semiorthogonal. We may extend $\bar A$ and $\bar B$ with further columns, which are normalized, orthogonal to the original columns and to each other to get orthogonal matrices, and also we may extend matrix $S$ by zero elements to size $m_A\times m_B$. Let us denote the respective extended matrices by $\tilde A$, $\tilde B$ and $\tilde S$. Then
\begin{equation}
M_{ij}=\sum_{k=1}^{m_A}\sum_{l=1}^{m_B}\tilde A_{ik}\tilde S_{kl}\tilde B_{lj}^T.
\label{eq:Bellsingf}
\end{equation}
The equation above is nothing else than a singular value decomposition of the Bell-matrix. The singular values are the diagonal elements of $\tilde S$, that is $\sqrt{m_Am_B}/d$ and zero. Therefore, $||g||_2=\sqrt{m_Am_B}/d$, and the upper bound for the quantum value is $m_Am_B/d$, which is achievable with the choice $\vec a_i=\vec A_i$ and $\vec b_i=\vec B_i$ (see Eq.~\ref{eq:Th1}). Therefore, Bell inequalities constructed this way from vectors obeying the constraint in the Observation are special cases of the ones discussed in Ref.~\cite{EppKamBru2}. Actually, Eq.~(\ref{eq:Bellsing}) is just the truncated singular value decomposition referred to in Theorem 1 in Ref.~\cite{EppKamBru2}, and the matrix denoted by $\alpha$ there, whose existence is required, is $\sqrt{m_A/d}\delta_{ij}$, which leads to the optimum measurement vectors given above, indeed.

In Ref.~\cite{EppKamBru2} the authors have also discussed modifications of the inequalities leaving the upper bound for the maximum quantum value achievable. One is twisting (Corollary 1i), which is nothing else for the present construction than changing the relative orientation of sets $\vec A$ and $\vec B$. The maximum quantum value remains the same, which is evident from the formula giving it, as we have already noted. As the local value depends on it, we will look for the optimum relative orientation minimizing the local value, consequently, maximizing the quantum violation. The other modification changes the singular values. If it is done appropriately as prescribed in Corollary 1ii, the bound for the maximum quantum value will change in a well-defined way while it remains achievable. We will use this modification only in a very special way and for a case when $\vec A_i=\vec B_i$. We subtract $\lambda$, the same number from all diagonal elements of $\tilde S$. It is allowed in this case if $\lambda<m_A/2d$. Then the modified Bell matrix will be: 
\begin{align}
M'_{ij}&=\sum_{k}^{m_A}\tilde A_{ik}(\tilde S_{kk}-\lambda)\tilde A^T_{kj}\nonumber\\
&=\sum_{k=1}^{d}\bar A_{ik}S_{kk}\bar A^T_{kj}-\lambda\sum_{k=1}^{m_A}\tilde A_{ik}\tilde A^T_{kj}=M_{ij}-\lambda\delta_{ij}.
\label{eq:modem}
\end{align}
Here we have used that $\tilde A_{ik}=\bar A_{ik}$ and $\tilde S_{kk}=S_{kk}$ if $k\le d$, and Eq.~(\ref{eq:Bellsing}). The modification of the Bell matrix is very simple, the same number is subtracted from each diagonal element. The maximum quantum value becomes $m_A^2/d-\lambda m_A$ and it can be achieved with measurement vectors $\vec a_i=\vec b_i=\vec A_i$, the same ones as without the modification. As it is easy to show that the local value decreases by the same $\lambda m_A$ amount than the quantum maximum until the optimal local strategy does not change, if the original inequality is violated, the modification increases the violation.

We have shown above that $\vec a_i=\vec A_i$ and $\vec b_j=\vec B_j$ is an optimal choice for the measurement vectors if both sets of construction vectors obey the constraint of the Observation. However this is not the only optimal choice. For example, the quantum value is obviously the same if $\vec a_i=\hat O\vec A_i$ and $\vec b_j=\hat O\vec B_j$, where $\hat O$ is an orthogonal transformation. There may be less trivial examples, it may even happen that measurement vectors in a lower dimensional subspace also lead to the maximum quantum value.

For the five three-dimensional and for the six four-dimensional Platonic solids the Observation may easily be verified explicitly. For more than four dimensional spaces there are only three kinds of Platonic solids, the cross polytope, the simplex and the hypercube (octahedron, tetrahedron and cube in three dimensions, respectively). In Appendix~\ref{specpropproof} we prove that the Observation is true for these bodies in all dimensions. We also prove that it is also true for all convex polygons, which are the Platonic solids of two dimensions. The 13 Archimedean solids in three dimensions also obey the property, as it can easily be checked.

The validity of the Observation is certainly connected to some symmetry properties, therefore, it must be valid for many other symmetric objects. The Observation involves $d(d+1)/2-1$ constraints. Finding the same number of independent constraints for a set of unit vectors which can be derived from the original ones is enough to show that the set has the property, with all of its consequences. Such constraints can be found in case of symmetric objects. For any $\hat O$ symmetry of $\hat V_i$ (that is when the set $\hat O\vec V_i$ is the same as $\hat V_i$), it is true that $\sum_{i=1}^m (\vec V_i\cdot \vec x)(\vec V_i\cdot \vec y)=\sum_{i=1}^m(\vec V_i\cdot\hat O^{-1}\vec x)(\vec V_i\cdot\hat O^{-1}\vec y)$. From Eq.~(\ref{eq:twodcontrhigd}) in Appendix~\ref{specpropproof} we also show that if the vectors have an $l$-fold symmetry for rotations in some plane, then for all vectors $\vec x$ in that plane the value of the sum $\sum_{i=1}^m (\vec V_i\cdot \vec x)(\vec V_i\cdot \vec x)$ are the same, and for all orthogonal pairs of vectors $\vec x$ and $\vec y$ in that plane $\sum_{i=1}^m(\vec V_i\cdot \vec y)(\vec V_i\cdot \vec x)=0$. If there are several such planes and possibly some other symmetries, it may be enough to ensure that all constraints are satisfied. 

It is easy to see that if the set $\vec V_i$ is centrally symmetric and obeys the constraints of the Observation, then the subset of $\vec V_i$ containing only one of each pairs pointing towards opposite directions also obeys the constraints (the scalar products of the columns are just one half of the scalar products corresponding to the original set).  

\section{The local bound}\label{Lbound}
In this paper we are dealing with Bell inequalities whose coefficients are given as scalar products of $d$-dimensional unit vectors from two sets. The considerations of this section do not require that the sets are normalized, which means that they are valid for any correlation-type bipartite Bell inequality, as the Bell coefficients can always be written as scalar products of two non-normalized sets of vectors using the singular value decomposition of the Bell matrix. The dimensionality of the vectors is equal to the rank of the Bell matrix. We further extend the validity of our analysis to generic non-correlation-type Bell inequalities. 

Normally, the computation time required for the exact calculation of the local bound grows exponentially with the number of measurement settings. With the method proposed in this section the dependence becomes polynomial whose degree is the rank of the Bell matrix. Therefore, if this rank is considerably lower than the number of settings on each side, a lot of computation effort can be saved. Unfortunately, if the rank is not low enough, the computation effort required still grows fast with the number of settings. Nevertheless, for low ranks and not very many settings the method makes exact calculations of local bounds affordable that would be intractable otherwise.  

The local bound for any correlation-type Bell inequality can be determined as:
\begin{equation}
L=\max\limits_{a_i,b_j=\pm 1}\sum_{i=1}^{m_A}\sum_{j=1}^{m_B}a_ib_j(\vec A_i\cdot\vec B_j),
\label{eq:localval}
\end{equation}
where $a_i$ and $b_j$ are the respective components of the local strategies of Alice and Bob, and $\vec A_i$ and $\vec B_j$ are not necessarily normalized Euclidean vectors. The evaluation of the expression would involve to check $2^{m_A+m_B}$ strategies. This can be reduced by noticing that for any strategy of Bob there is an optimal strategy for Alice, namely: 
\begin{equation}
a_i={\text{sgn}}\left(\vec A_i\cdot\sum_{j=1}^{m_B}b_j\vec B_j\right).
\label{eq:optai}
\end{equation}
Using this strategy for Alice, the local bound becomes:
\begin{equation}
L=\max\limits_{b_j=\pm 1}\sum_{i=1}^{m_A} \sum_{j=1}^{m_B}|\vec A_i\cdot b_j\vec B_j|.
\label{eq:localval1}
\end{equation}
Taking into account that strategies with components $b_j$ and $-b_j$ give the same value, the evaluation of $2^{m_B-1}$ expressions is sufficient (for example, one can consider only strategies with $b_1=+1$). Obviously, if $m_A<m_B$, one can use an analogous formula and end up with $2^{m_A-1}$ evaluations. However, the present construction of the Bell coefficients may lead to further and very significant reduction of the number of evaluations. Instead of an exponential dependence on the number of settings we get a polynomial dependence whose degree is $d-1$. 

Furthermore, our result for bipartite correlation-type inequalities can be extended to generic bipartite two-outcome inequalities. Indeed, in this case we can write the Bell inequality
\begin{equation}
\sum_{i=1}^{m_A}M^A_i \langle{a_i}\rangle+\sum_{j=1}^{m_B}M^B_j \langle{b_j}\rangle+\sum_{i=1}^{m_A}\sum_{j=1}^{m_B}M_{ij}\langle{a_ib_j}\rangle\leq L.
\label{eq:margbell}
\end{equation}
We show that the calculation of the local bound above can be traced back to the calculation of the local bound of a correlation-type Bell inequality~(\ref{eq:corbell}). In particular, let us define the following $(m_A+1)\times(m_B+1)$-setting correlation-type inequality:
\begin{align}
&M_{m_A+1,m_B+1}\langle a_{m_A+1}b_{m_B+1}\rangle+\sum_{i=1}^{m_A}M^A_i \langle{a_i}b_{m_B+1}\rangle\nonumber\\
&+\sum_{j=1}^{m_B}M^B_j \langle{a_{m_A+1}b_j}\rangle+\sum_{i=1}^{m_A}\sum_{j=1}^{m_B}M_{ij}\langle{a_ib_j}\rangle\leq L'.
\label{eq:cor2bell}
\end{align}
Let us further choose the constant
\begin{equation}
M_{m_A+1,m_B+1} = \sum_{i=1}^{m_A} |M^A_i| + \sum_{j=1}^{m_B}|M^B_j| + \sum_{i=1}^{m_A}\sum_{j=1}^{m_B}|M_{ij}|. 
\end{equation}
In this case, one can show that the local bound $L'$ above is obtained with a deterministic strategy, where $a_{m_A+1}b_{m_B+1}=1$. Then the local bound $L$ in (\ref{eq:margbell}) can be obtained by a simple subtraction from the local bound $L'$ in (\ref{eq:cor2bell}): $L=L'-M_{m_A+1,m_B+1}$. On the other hand, the Bell matrix in Eq.~(\ref{eq:cor2bell}) can have only one more rank than the Bell matrix in Eq.~(\ref{eq:margbell}). This follows from the definition of the matrix rank given by the linearly independent number of rows or columns of the matrix. Thus, we proved that the local bound $L$ of a generic two-outcome bipartite Bell inequality~(\ref{eq:margbell}) can be reduced to the calculation of the local bound $L'$ of the correlation-type Bell inequality~(\ref{eq:cor2bell}). Using the following algorithm to compute the local bound of a correlation-type Bell inequality with small matrix rank $M$, however, we obtain an efficient algorithm for computating the local bound of any two-outcome bipartite Bell inequality when the matrix rank $M$ is much lower than the number of settings $m_A$ and $m_B$.

Let us call strategy $a_i$ a geometrical strategy for Alice if there exists some vector $\vec q$ such that $a_i={\text{sgn}}(\vec A_i\cdot \vec q)$. Looking at Eq.~(\ref{eq:optai}) it is clear that all strategies of Alice that are potentially optimal are geometrical ones. From the analogous expression giving the optimal $b_j$ for $a_i$ follows that the same is true for the potentially optimal strategies of Bob. Therefore, it is enough to check only the geometrical strategies when calculating the local bound. From Eq.~(\ref{eq:localval1}) it follows that applying a geometrical strategy of Bob means multiplying $\vec B_i$ by $+1$ or $-1$ depending on which side of a hyperplane (the normal plane of $\vec q$) it is, such that $b_i\vec B_i$ are all in the same side.

Now we give a recipe to determine all geometrical strategies for a set of vectors $\vec A_i$. We may take $\vec q$ to be a unit vector, so that it may be thought of as a point on the surface of the unit sphere. It is clear that a small change in $\vec q$ usually will not change the geometrical strategy it defines: there is a whole region on the surface of the unit sphere giving the same geometrical strategy. If we move around on the surface and cross the normal hyperplanes of one of the $\vec A_i$ vectors, than $a_i$ will change sign. Until such does not happen, we remain in the same region. Therefore, the borders of the regions corresponding to geometrical strategies are the intersections of the normal hyperplanes of $\vec A_i$ with the unit sphere, which are great (hyper)circles of the (hyper)sphere.

The question is how to find all regions, or at least the geometrical strategies they define. In two dimensions the problem is easy. The unit sphere is a circle and its intersections with the normal planes (now just lines), that is the borders of the regions are points which are easy to find. There are $2m_A$ such points, consequently the same number of regions and geometrical strategies (if no two vectors point to the same or opposite directions).

Let us consider the three-dimensional case, which is not difficult to visualize and which can be generalized to higher dimensions. First let us discuss the number of geometrical strategies, that is the number of regions on the surface of the sphere whose borders are the great circles defined by the normal planes of $\vec A_i$. If there is only one vector, then the one great circle divides the surface into two regions. For two vectors, if they do not point to the same or to the opposite directions, the number of regions is four. If we add a third vector such that only two great circles cross each other in all intersections, that is there are no multiple crossings, the number of regions will be doubled again to eight.  There is no gain so far, all strategies are geometrical ones. However, when a fourth vector is added, the number of regions grows only by six instead of eight, and the $(i+1)$th vector will increase this number only by $2i$ instead of $2^i$. The reason is simple. The new great circle crosses each existing one twice. Between each consequtive crossings one region is cut into two parts. Since there are $2i$ new crossings, the number of regions grows by this amount. It is easy to prove by induction that if there are $m_A$ vectors, then the number of regions is $m_A(m_A-1)+2$. If there are multiple crossings, the number of regions will be smaller: if two crossings occur in the same place, there is no region cut between them. For symmetric arrangements, like the Platonic bodies, multiple crossings occur. Although a smaller number of geometric strategies is advantageous in itself, unfortunately the determination of them is more difficult, as we will see later.
 
Our aim is to determine all geometrical strategies. To do that let us concentrate on the intersections, which are the corners of the regions. Each intersection is in a direction orthogonal to two vectors. All intersections can be found in the directions of vectors $\pm\vec A_k\times\vec A_l$ taken for all pairs ($k,l=1,\dots,m_A$). We have supposed that there are no pairs of vectors pointing towards opposite directions. Let us also suppose that there are no multiple crossings. Then each intersection is on the border of four regions. For the four strategies corresponding to these regions each $a_i$ for $i\neq k; i\neq l$ are the same for all four regions and they are given as the sign of the scalar product of $\vec A_i$ with the vector pointing towards the intersection (these are well defined in case of no multiple intersections), while for $i=k$ or $i=l$, for which this scalar product is zero, the four variations of $+1$ and $-1$ has to be taken. This way we get all geometrical strategies, but in multiple copies, each strategy will occur in our list as many times as the number of corners of its region. Therefore, the last step is to get rid of the multiple copies.

If $a_i$ represents a geometrical strategy so does $-a_i$. One can save some effort when applying the algorithm above by looking only for one from each such pair. Therefore, it is better to take only one of the intersections from the pair in opposite directions (say $+\vec A_k\times\vec A_l$), and to reverse all signs whenever a strategy derived has $a_1=-1$.

The algorithm may be extended to dimensions $d>3$. Now the corners of the regions, that is the points to find are at the intersections of $d-1$ hyperplanes with the hypersphere. They are in directions which are orthogonal to $d-1$ vectors. This is towards the null space of the matrix given by those $d-1$ vectors, provided the null space is one-dimensional (if not, those vectors do not define a corner). One may find all corners by taking all combinations of $d-1$ vectors out of the $m_A$ ones. Each such point now is on the border of $2^{d-1}$ regions. The strategies corresponding to these regions can be determined similarly to the three-dimensional case. For $d=4$ the number of geometrical strategies is  $m_A(m_A-1)(m_A-2)/3+2m_A$. We note that if $m_A\leq d$ all strategies are geometrical ones if $\vec A_i$ are linearly independent.

So far, we supposed that there are no pairs of vectors pointing towards opposite directions and no multiple crossings, meaning that no more than the minimum number of great hypercircles to define a point intersect in each point. For randomly chosen $\vec A_i$ this is true. Unfortunately in the case of Platonic solids, and when $\vec A_i$ shows some symmetries, it is not the case. If $\vec A_i=-\vec A_j$ for some $i,j$ then in all geometric strategies it is obviously true that $a_i=-a_j$. Therefore, we may eliminate one of them and reduce the number of settings $m_A$ by one by dropping one of the vectors and doubling the length of the other one. We note that in this part we have not used that $A_i$ are normalized, all considerations are true even if their lengths are different. For $d>2$ all Platonic bodies are centrally symmetric except for the simplices, therefore one half of the vertices can be eliminated. We may even leave the vectors normalized and just multiply the local bound by two at the end.

The most obvious way to treat the problem of multiple crossings is to take all variations of $+1$ and $-1$ for the undetermined strategy components, that is if $\nu$ great hypercycles intersect, create $2^\nu$ strategies. However, this way we would create spurious strategies. Let us look at the three-dimensional case. If there is a $\nu$-fold intersection, we would create $2^{\nu}$ strategies, while the intersection is on the border of only $2\nu$ regions. This recipe may even be intractable. For the four dimensional Platonic body that has the most vertices (the dodecaplex) even 30-fold intersections occur. It would probably be possible to work out an algorithm to identify the truely existing regions, but it does not seem obvious. We have chosen a different approach. We can dissolve the degeneracies of the intersections by applying small perturbations to vectors $\vec A_i$. This method will also introduce spurious strategies, but not as many as the one discussed above. If the perturbations are small enough, the desired strategies will always be found. The number of all strategies created this way, including the spurious ones, will be the same as we would get for random $\vec A_i$. The spurious strategies we get depend on the perturbation applied. Therefore, we may get rid of them by repeating the calculation with different perturbations several times and keeping the strategies present in the results of all runs.

For the dodecaplex, the four dimensional Platonic solid that has 600 vertices, the total number of strategies to consider is $2^{599}$, which can be reduced to $2^{299}$ by eliminating half of the vertices, as explained earlier. For the same object the number of geometric strategies to be checked is just 1787760.

For further considerations it is useful to introduce the notion of strategy vectors that we define as $\sum_{i=1}^{m_A}a_i\vec A_i$ and $\sum_{j=1}^{m_B}b_i\vec B_i$ for Alice and Bob, respectively, where $a_i$ and $b_j$ are geometric strategies. The local bound in Eq.~(\ref{eq:localval}) may be rewritten as:
\begin{equation}
L=\max\limits_{a_i,b_j {\text{geom. strat.}}}\left(\sum_{i=1}^{m_A}a_i\vec A_i\right)\cdot\left(\sum_{j=1}^{m_B}b_j\vec B_j\right).
\label{eq:localvalst}
\end{equation}
This means that the local value is the largest of the scalar products of Alice's and Bob's strategy vectors. If we vary the relative orientation of sets $\vec A_i$ and $\vec B_j$ by applying orthogonal transformations to one of them, when the longest strategy vectors overlap, the local value is just the product of their length. This is obviously the maximum value of $L$ in any relative orientation. For symmetric arrangements, like in the case of Platonic bodies, it may happen that there are many strategy vectors having the same maximal length. In this case, if the dimensionality of the space is not large, there will be no relative orientation with a local bound significantly smaller than this maximum value, since there will always be a pair of such vectors with a small angle between them. 

\section{The two dimensional case}\label{Twodim}
The Platonic solids of two dimensions are the regular convex polygons. There is an infinite number of them. Let $\vec A_i$ point towards the vertices of an $m_A$-sided regular convex polygon. Then the lengths of all strategy vectors will be the same. If $m_A$ is even, this value is $l_{m_A}=2\sum_{k=1}^{m_A/2}\sin[(2k-1)\pi/m_A]$, and the vectors point towards the midpoints of the sides if $m_A$ is divisible by four, and towards the vertices, if not. If $m_A$ is odd, there are $2m_A$ strategy vectors the same as for the $2m_A$-sided polygon, except that their length is halved. Half of them point towards the vertices and half of them towards the midpoints of the sides. The strategy vectors have length $l_{m_A}=\sum_{k=1}^{m_A}\sin[(2k-1)\pi/2m_A]$.

\begin{figure}[t!]
\begin{center}
\includegraphics[width=7.cm]{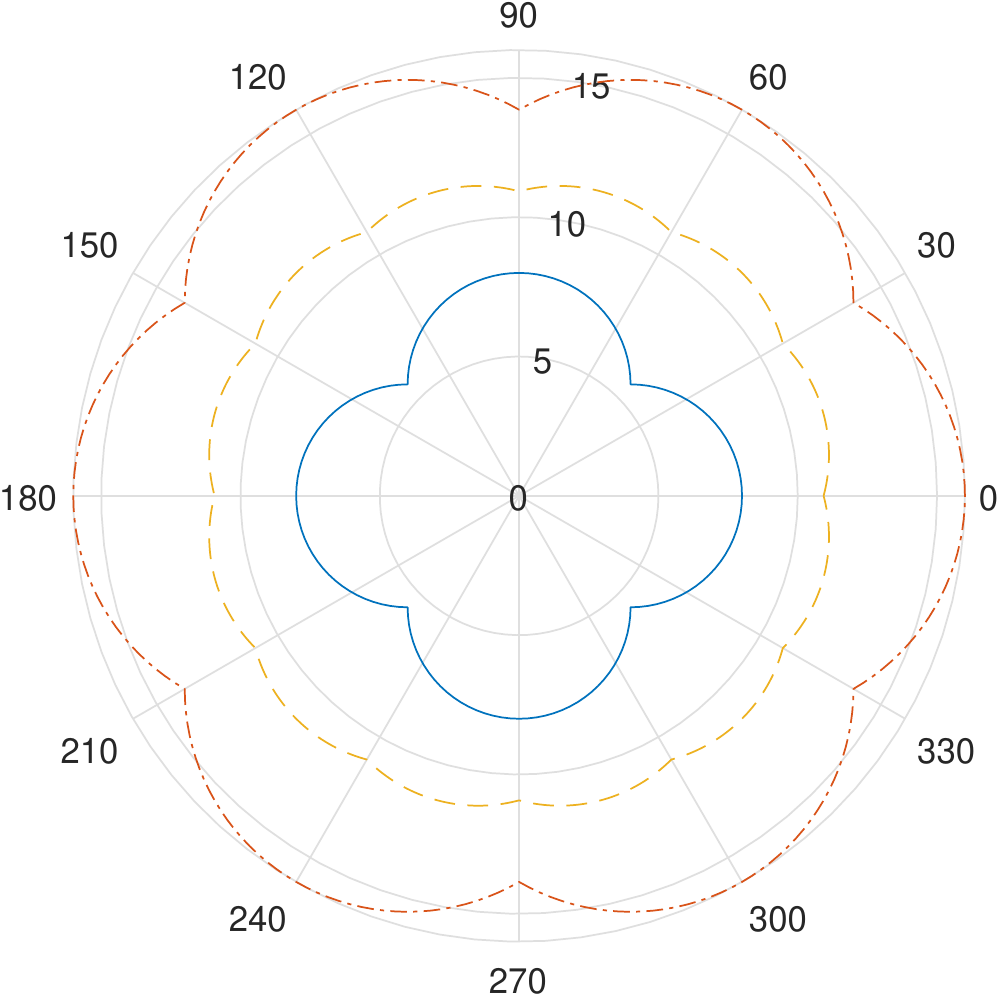}
\end{center}
\caption{Dependence of the local bound on the relative orientation of the two regular polygons used to construct the Bell inequality. The solid, the dash-dotted and the dashed line correspond to two squares, two hexagons and a square and a hexagon, respectively. At zero angle (at least) two vertices align.\label{fig_2d}}
\end{figure}

If there are identically oriented identical polygons on both sides ($\vec A_i=\vec B_i$), the local bound of the Bell inequality constructed is the square of the length of the  strategy vector, while the quantum bound with any relative orientation is $m_A^2/2$. Therefore, for two squares the local bound is the same as the quantum bound, namely 8, so there is no violation. However, if we rotate one of the squares by $\pi/4$ the local bound is multiplied by $1/\sqrt{2}$ (the cosine of the largest angle between pairs of strategy vecors from the two sides), therefore $Q/L$ becomes $\sqrt{2}$ (the Bell inequality is like four CHSH ones~\cite{CHSH} multiplied by $1/\sqrt{2}$ or $-1/\sqrt{2}$). In case of two regular hexagons the maximum local bound is $16$ (identical orientation) and the minimum one is $8\sqrt{3}=13.856406$ ($\pi/6$ relative angle). As the quantum bound is 18, the minimum $Q/L$ ratio is $9/8=1.125$, while the maximum one is $9/(4\sqrt{3})=1.2990381$. The values are the same for  regular triangles. If there are two identical regular even-sided polygons, with one of them rotated by angle $\varphi$, then for $-\pi/m\leq\varphi\leq\pi/m$ (where $m=m_A=m_B$) the local bound is $l_{m}^2\cos\varphi$. At the ends of the interval the function is minimal, and the pattern repeated $m$-times around the circle. The ratio of the minimal and the maximal local bound is $\cos(\pi/m)$. In Fig.~\ref{fig_2d} we show the $\varphi$-dependence of the local bound for two squares and for two hexagons.

The behaviour is qualitatively the same for two non-identical even-sided polygons, too, but the periodicity of the function is $2\pi/\text{lcm}(m_A,m_B)$, where lcm denotes the least common multiple. In Fig.~\ref{fig_2d} we also show the result for a square on one side and a hexagon on the other side. The length of the period can be determined by finding out that starting from a position when two strategy vectors overlap, at least how much we have to rotate one of the polygons such that two other vectors overlap again. Let the directions of the strategy vectors in the starting position be given by angles $2\pi\mu/m_A$ ($\mu=0,\cdots,m_A-1$) and $2\pi\nu/m_B$ ($\nu=0,\cdots,m_B-1$). Then the angles between the pairs are
\begin{equation}
2\pi\left|\frac{\mu}{m_A}-\frac{\nu}{m_B}\right|=2\pi\frac{|\mu m'_B-\nu m'_A|}{{\text{lcm}}(m_A,m_B)},
\label{eq:angles}
\end{equation}
where $m'_A=m_A/{\text{gcd}}(m_A,m_B)$ and $m'_B=m_B/{\text{gcd}}(m_A,m_B)$, and gcd refers to the greatest common divisor. As $m'_A$ és $m'_B$ are relative primes, the minimal nonzero value of the numerator on the right hand side is one, therefore, the smallest nonzero value of the expression is $2\pi/\text{lcm}(m_A,m_B)$, indeed, which is the period of the function we are looking for. The ratio between the minimal and the maximal local bound is $\cos(\pi/\text{lcm}(m_A,m_B))$. For odd-sided polygons the behaviour is the same as for the ones with twice as many sides, only the local bound and the quantum value has to be divided by two.

For large $m_A$ the length of the strategy vectors $l(m_A)$ converges to $2m_A/\pi$, because $l_{m_A}/m_A$ approaches the Riemann-sum of $\int_{x=1}^{\pi}\sin(x)dx/\pi$. Therefore, $Q/L$ converges to $\pi^2/8=1.2337006$ if both $m_A$ and $m_B$ tend to infinity. This is obviously true for any relative orientations of the polygons. This value recovers the $Q/L$ ratio of the inequality with a continuum infinite number of settings corresponding to the special case $n=2$ in Refs.~\cite{vp09,briet11}.

\section{Local strategies for $d\geq 3$}\label{Locstratdge3}

\begin{table*}[t]
\begin{center}
\begin{tabular}{ |c|c|c|c|c|c|c| } 
\hline
&$m$&cs&$n_{\rm{sv}}$&$n_{\rm{svg}}$&$n_{\rm{svx}}$&$m/\sqrt{3}/l_x$\\
 \hline
 tetrahedron&4&n&14&2&6&1.0000000\\
 \hline
 octahedron&6&y&8&1&8&1.0000000\\
 \hline
 cube&8&y&14&2&6&1.0000000\\
 \hline
 icosahedron&12&y&32&2&12&1.0704663\\
 \hline
 dodecahedron&20&y&92&3&20&1.1026409\\
 \hline
 truncated tetrahedron&12&n&134&5&8&1.1055416\\
 \hline
 cuboctahedron&12&y&24&1&24&1.0954451\\
 \hline
 truncated cube&24&y&134&6&6&1.0572365\\
 \hline
 truncated octahedron&24&y&116&4&12&1.0954451\\
 \hline
 rhombicuboctahedron&24&y&134&4&6&1.0978539\\
 \hline
 truncated cuboctahedron&48&y&554&16&6&1.1084962\\
 \hline
 snub cube&24&n&482&15&6&1.1074356\\
 \hline
 icosidodecahedron&30&y&120&1&120&1.1387895\\
 \hline
 truncated dodecahedron&60&y&872&12&20&1.1240927\\
 \hline
 truncated icosahedron&60&y&872&12&12&1.1397231\\
 \hline
 rhombicosidodecahedron&60&y&872&11&20&1.1414557\\
 \hline
 truncated icosidodecahedron&120&y&3542&35&30&1.1387895\\
 \hline
 snub dodecahedron&60&n&3422&37&30&1.1401438\\
\hline
\end{tabular}
\caption{Properties of the strategy vectors for the three dimensional Platonic and Archimedean solids. The number of vertices is $m$, cs shows if the solid is centrally symmetric, $n_{\rm{sv}}$ is the number of strategy vectors, $n_{\rm{svg}}$ is the number of groups of strategy vectors containing vectors of the same lengths, $n_{\rm{svx}}$ is the size of the group containing the longest strategy vectors of $l_x$ length. The $Q/L$ value for a Bell inequality constructed from two solids in a relative orientation such that two of their longest strategy vectors align is the product of the corresponding $m/\sqrt{3}/l_x$ values shown in the last column. For any different relative orientation $Q/L$ is larger. \label{table:3dsolids}}
\end{center}
\end{table*}

First we discuss the three Platonic solids existing in all dimensions, the cross polytope, the simplex and the hypercube. We note that in the present context in $d=3$ the simplex (tetrahedron) and the cube are equivalent: one can get the cube from the tetrahedron by reflecting its vertices to the  midpoint and adding the resulting points as new vertices. There is a one-to-one correspondence between the geometrical strategies for the two bodies, in the cube the component corresponding to the reflected vertex is minus one times the value for its mirror image, therefore, all strategy vectors will be the same only with double lengths. If used as construction vectors on one side for a Bell inequality, both the classical and the quantum bound should simply be multiplied by two. For $d>3$ there are no such equivalences.

From the present pont of view the simplest body is the {\bf cross polytope}. As it has been told before, it has $2d$ vertices, in the most convenient orientation they are the basis vectors with both signs, that is $\vec A_i=\vec e_i$ and $\vec A_{i+d}=-\vec e_i$, ($i=1,\dots,d$). The geometrical strategy defined by vector $\vec q$ is $a_i={\text{sgn}}(\vec A_i\cdot \vec q)$, which is $a_i=-a_{i+d}={\text{sgn}}(q_i)$. The strategy vector $\sum_{i=1}^{2d}a_i\vec A_i=2\sum_{i=1}^d\text{sgn}(q_i)\vec e_i$. All $\vec q$ whose every component has the same sign give the same strategy vector. Therefore, each strategy vector can be written as $2\sum_{i=1}^d \alpha_i\vec e_i$, where $\alpha_i$ takes the value of $+1$ or $-1$, and all such combinations are to be taken to get all strategy vectors. There are $2^d$ vectors, they point towards the vertices of a hypercube and have the same length of $2\sqrt{d}=m_A/\sqrt{d}$. As the quantum bound for a Bell inequality constructed from two identical bodies is $m_A^2/d$, and the local bound is the square  of the length of the longest strategy vector if the orientation is the same in both sides, for two such cross polytopes there is no violation.

A {\bf simplex} has $d+1$ vertices. Each pair of vertices has the same scalar product $\vec A_i\cdot\vec A_j=-1/d$, $i\neq j$. When constructing all strategies, we now consider not the $\vec q_i$ vectors but their normal hyperplanes. If such a plane cuts the $d$-dimensional space, any $1\leq k\leq d$ of the $d+1$ vertices may get into one side and the rest into the other side. Such a cut gives a strategy with $k$ plus one and $d+1-k$ minus one components (from all strategies only the two strategies corresponding to all signs equal are not geometrical ones). For a given $k$ there are $\binom{d+1}{k}$ possibilities, corresponding to the same number of strategies and strategy vectors of lengths $2\sqrt{k(d+1-k)/d}$. The reason is the following. Due to the symmetry $\sum_{i=1}^{d+1}\vec A_i=0$, therefore, the sum of the $k$ of them multiplied by one is the same as the sum of the $d+1-k$ of them multiplied by minus one. As $(\sum_{i=1}^{k}\vec A_i)\cdot(\sum_{j=1}^{k}\vec A_j)=k-k(k-1)/d$ (any $k$ vectors could have been used as the ones on the plus side), and the length of the strategy vector is twice the square root of this value, what we get is just what we have given above, indeed. The longest strategy vectors correspond to the most even distribution of vectors, that is $k=(d+1)/2$ if $d$ is odd, and $k=d/2$ and $k=d/2+1$ if $d$ is even, that is the maximum length is $(d+1)/\sqrt{d}$ and $\sqrt{d+2}$ for odd and even $d$, respectively. When $d$ is odd (similarly to the cross polytope in all dimensions), the Bell inequality constructed using two simplices of the same orientation can not be violated (both the local and the quantum bound is $(d+1)^2/d$).

The $d$-dimensional {\bf hypercube}, the $d$-cube has $m_A=2^d$ vertices $\vec A_i$, whose each coordinate $A_{ij}$ in the most convenient orientation is either $+1/\sqrt{d}$ or $-1/\sqrt{d}$, such that all variations of the signs are covered. If $\hat T$ is an orthogonal transformation that permutes the coordinates and change the sign of some of them, the set $\hat T\vec A_i$ contains the same vectors as $\vec A_i$, only in a different order. From this it follows that if $\vec S$ is a strategy vector, so is $\hat T\vec S$. If $\vec S$ is a strategy vector there exists $\vec q$ such that $\vec S=\sum_{i=1}^{2^d}\text{sgn}(\vec q\cdot\vec A_i)\vec A_i$. Then the strategy vector belonging to $\hat T\vec q$ is $\sum_{i=1}^{2^d}\text{sgn}(\hat T\vec q\cdot\vec A_i)\vec A_i=\sum_{i=1}^{2^d}\text{sgn}(\hat T\vec q\cdot\hat T\vec A_i)\hat T\vec A_i=\sum_{i=1}^{2^d}\text{sgn}(\vec q\cdot\vec A_i)\hat T\vec A_i=\hat T\vec S$. The first equality follows from the fact that sets $\vec A_i$ and $\hat T\vec A_i$ are the same, while the second one from the orthogonality of $\hat T$. From this it follows that the strategy vectors of the $d$-cube belong to families, where the members of a family can be derived by taking all permutations and all variations of the signs of the coordinates of one of its members. For $d=3$ there are two families that may be represented by $(0,0,8)/\sqrt{3}$  and $(4,4,4)/\sqrt{3}$ the members of the families point towards the midpoints of the faces and the vertices, respectively. There are altogether $6+8=14$ strategy vectors, each corresponding to a geometrical strategy. For $d=4$ there are three families represented by $(0,0,0,8)$, $(0,4,4,4)$ and $(2,2,2,6)$. The total number of strategy vectors is $8+32+64=104$. For $d=5$ the number of families is $7$, while the number of vectors is $10+80+320+160+960+320+32=1882$. The number of families for $d=6,7$ and $8$ are $21,135$ and $2470$, respectively, while for $d=9$ it is at least $175428$. Unfortunately, we have not found a systematic way to get formulae for all dimensions. However, it is true for all dimensions that the longest strategy vectors point towards the midpoints of the facets, they are $2^d\vec e_i/\sqrt{d}=\sum_{j=1}^{2^d}(\vec e_i\cdot\vec A_j)\vec A_j$ (all components but the $i$th one add up to zero), which shows that they are strategy vectors. As their length is $2^d/\sqrt{d}=m_A/\sqrt{d}$, the Bell inequality constructed from two $d$-cubes of the same orientation can not be violated. These strategy vectors are indeed the longest, since the local bound cannot be greater than the quantum bound.

It is true for all cases we have considered so far except for simplices in even dimensions that the length of the longest strategy vectors is $m_A/\sqrt{d}$. Therefore, if a Bell inequality is constructed from any two such objects, there exist relative orientations (when longest strategy vectors are aligned), when the inequality cannot be violated. If one of the bodies is a simplex in an even dimension, $Q/L$ is $(d+1)/\sqrt{d(d+2)}$ if the orientation is such that the local bound is maximal, while it is the square of this number if there are simplices in both sides.

For $d=3$ there are two further Platonic solids, the icosahedron and the dodecahedron with 12 and 20 vertices, respectively, while for $d=4$ we have the octaplex, the tetraplex and the dodecaplex with 24, 120 and 600 vertices, respectively. We have calculated the geometrical strategies and the strategy vectors using the algorithm presented in Section~\ref{Lbound}. All these bodies are centrally symmetric, therefore when we applied the algorithm we eliminated one half of their vertices.

In three dimensions for the {\bf icosahedron} there are 32 geometrical strategies and strategy vectors. The longest ones of length $2\sqrt{6+2\sqrt{5}}=6.4721360$ point towards the 12 vertices, while the shorter ones of lengths $2\sqrt{6+6/\sqrt{5}}=5.8934817$ point towards the middle of the 20 faces (as the coordinates of the vertices are known analytically, if we know a geometrical strategy, we may derive the corresponding strategy vector analytically). If the Bell inequality is constructed from two identical sets corresponding to the icosahedron, then the local bound is $41.888544$, and as the maximum quantum value is $12^2/3=48$, $Q/L=1.1458980$. For the {\bf dodecahedron} there are $92$ geometric strategies. The length of the $20$ longest strategy vectors is $10.472136$, and there are also $60$ and $12$ further vectors of lengths $9.8863510$ and $9.8224695$. The analytical forms are quite involved. For two identically oriented dodecahedrons $L=109.66563$ and $Q/L=1.2158169$. Results for the strategy vectors for all three dimensional Platonic solids and also for all Archimedean solids are summarized in Table~\ref{table:3dsolids}.

In four dimensions there are $192$ geometrical strategies for the {\bf octaplex}, and all strategy vectors have the same length of $4\sqrt{7}=10.583005$. Therefore, for two identical bodies $L=112$ and $Q/L=144/112=1.2857143$. For the {\bf tetraplex} the $14400$ strategy vectors also have equal lengths of $51.146605$, therefore for two such objects $L=2615.9752$ and $Q/L=1.3761599$. For the {\bf dodecaplex} there are $3575520$ strategy vectors. Surprisingly, they can be arranged into $201$ groups of different lengths. One of the smallest group of $1440$ members is the one containing the longest vectors of length $255.71725$. Consequently, for two identical bodies $L=6539.1314$ and $Q/L=1.3763296$. The shortest strategy vectors belong to one of the largest groups of $129600$ members, and their length is $254.17151$. Although there are many groups, the difference between the longest and shortest vectors is just $0.6\%$.

As we have said, when two objects are oriented such that their longest strategy vectors point into the same direction, the corresponding Bell inequality has the maximum local bound. This can be reduced by changing the relative orientation. If both objects obey the constraint given in the Observation, reorientation will not change the quantum bound, therefore $Q/L$ will increase, like we have already discussed in the two-dimensional cases. We explore in the next section how much improvement we can achieve this way for $d>2$.


\section{Minimal local bounds for $d\geq 3$}\label{MinLocsdge3}

To find relative orientations where the local bound is minimal, consequently $Q/L$ is maximal, we have used the Nelder-Mead method starting from many random relative orientations. The transformation has been parameterized by the $d(d-1)/2$ generalized Tait–Bryan angles. Unfortunately, the minimal values found this way are not necessarily the global minima, it may happen that is some cases there exists relative orientations with lower local bounds, but we have missed them. Whenever the numerical result has shown some recognizable  analytical pattern, we have tried to find the analytical form (either for the transformation or for the Bell inequality itself), and give $L$ analytically, or as accurately as possible.

The simplest body from our point of view is the cross polytope, a centrally symmetric object. If we eliminate half of its vertices in the most obvious way, the remaining ones are just the basis vectors $\vec e_i$, ($i=1,\dots,d$). If these are the vectors on one side to construct the Bell inequality, then the Bell coefficients will be nothing else than the vector coordinates on the other side.

{\bf Cross polytope--cross polytope}. Let another such halved cross polytope be on the other side, but reoriented by an orthogonal transformation. An optimal transformation we found consists of rotations in the $1-2$ the $3-4$, and so on planes by $\pi/4$ angle. If $d$ is odd, the $d$th coordinate remains unchanged. If $d$ is even, we get the sum of $d/2$ independent CHSH inequalities multiplied by $1/\sqrt{2}$. Therefore, the local bound is $d/\sqrt{2}$, while the quantum maximum is $d$, as it should be, giving $Q/L=\sqrt{2}$. For odd $d$ there are $(d-1)/2$ independent inequalities proportional to CHSH, and there is one additional nonzero Bell coefficient, namely $M_{dd}=1$. Then $L=(d-1)/\sqrt{2}+1$, $Q=d$ and $Q/L=\sqrt{2}/[1+(\sqrt{2}-1)/d]$. For two complete cross polytopes $L$ and $Q$ should be multiplied by four.

{\bf Cross polytope--hypercube}. Let there be a hypercube on Bob's side. When $d=3$ an optimal transformation we found was a rotation around the third axis (that is in the $1-2$ plane) by $\pi/3$, which leads to $L=8+8/\sqrt{3}=12.618802$ and $Q/L=1.2679492$. For $d=5$ the same transformation turned out to be optimal, namely a rotation in the $1-2$ plane by $\pi/3$, which gives $L=50.789699$ and $Q/L=1.2600980$. It is possible that this transformation is optimal in all odd dimensions if there is a cross polytope on one side and a hypercube on the other one. The local bound can be analytically calculated in this case, the result is $L=4\binom{d-1}{(d-1)/2}(\sqrt{3}+d-2)/\sqrt{d}$, and $Q/L=2^{d+1}/L$. The calculation can be done for even $d$ as well, but in that case such transformation is never optimal. For $d=4$ the optimal transformation is the same as in the case of two cross polytopes: $\pi/4$ rotations in the $1-2$ and in the $3-4$ planes, and then we obtain $L=32/\sqrt{2}$ and $Q/L=\sqrt{2}$. If we eliminate appropriately one half of vertices of both bodies, the Bell inequality we get will be like the sum of four CHSH ones divided by $\sqrt{2}$. Although they are not independent, there exist strategies and measurement settings that optimize all of them, therefore, $Q/L=\sqrt{2}$, indeed. Surprisingly, the analogous transformation is inferior for $d=6$. There the best we could find was to apply no rotation at all. Then $L=40\sqrt{6}=97.979590$, and $Q/L=1.3063945$. The local bound with no reorientation can be calculated for any $d$ analytically, the result is $L=2\sqrt{d}\binom{d}{d/2}$ for even $d$. It is possible that this is the best we can do for $d\geq 6$ even values, but we do not know. Actually, for large $d$ the formula approaches $\sqrt{2/\pi}2^{d+1}$, consequently $Q/L$ converges to $\sqrt{\pi/2}$. For odd $d$ what we get without reorientations is worse than what we obtained with the $\pi/3$ rotation in a single plane, as discussed above. 

\begin{table*}[t]
\begin{center}
\resizebox{17.00cm}{!}{
\begin{tabular}{ |c|c|c|c|c|c| } 
\hline
3D Platonic solids & tetrahedron & octahedron & cube & icosahedron & dodecahedron \\
 &\hphantom{dodecahedron}&\hphantom{dodecahedron}&\hphantom{dodecahedron}&\hphantom{dodecahedron}&\hphantom{dodecahedron}\\
 \hline
 tetrahedron&1.0000000&1.0000000&1.0000000&1.0704663&1.1026409\\
 $m=4$& 1.2360680&1.2679492&1.2360680&1.2584081&1.1803399\\
 \hline
 octahedron &&1.0000000&1.0000000&1.0704663&1.1026409\\
 $m=6$&&1.2426407&1.2679492&1.1969160&1.1755012\\
 \hline
 cube &&&1.0000000&1.0704663&1.1026409\\
 $m=8$&&&1.2360680&1.2584081&1.1803399\\
 \hline
 icosahedron &&&&1.1458980&1.1803399\\
 $m=12$&&&&1.2811529&1.2734640\\
 \hline
 dodecahedron &&&&&1.2158169\\
 $m=20$&&&&&1.2901781\\
\hline
\end{tabular}
}
\caption{Maximum quantum values divided by local bounds $Q/L$ for Bell inequalities constructed from unit vectors pointing towards the vertices of three-dimensional (3D) Platonic solids. The upper and the lower numbers correspond to relative orientations giving the maximal and the minimal local bounds, respectively. In the first column the number of vertices for the solids is also shown.\label{table:d3}}
\end{center}
\end{table*}

\begin{table*}[t]
\begin{center}
\resizebox{17.00cm}{!}{
\begin{tabular}{ |c|c|c|c|c|c|c| } 
\hline
 4D Platonic solids & pentacope & hexdecahedroid & tesseract & octaplex & tetraplex& dodecaplex \\
Orthographics & \begin{minipage}{0.6in}\includegraphics[width=\textwidth]{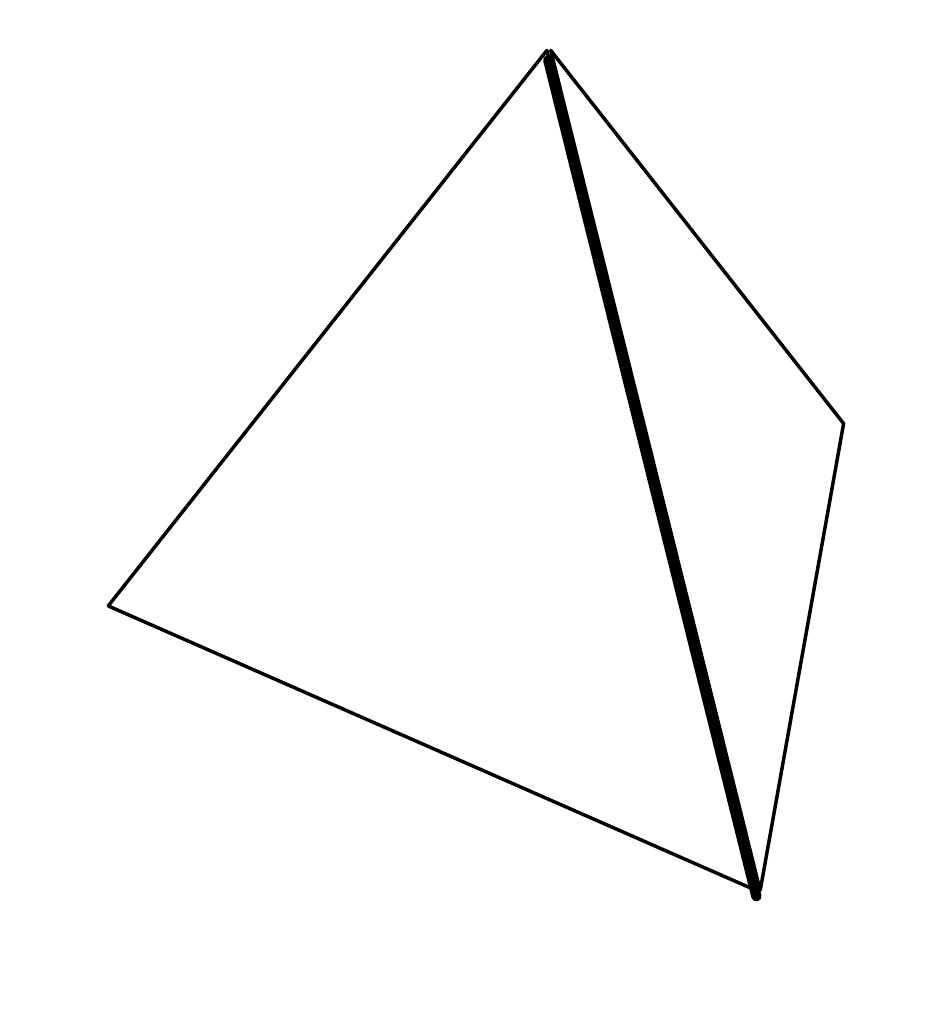}\end{minipage} & \begin{minipage}{0.6in}\includegraphics[width=\textwidth]{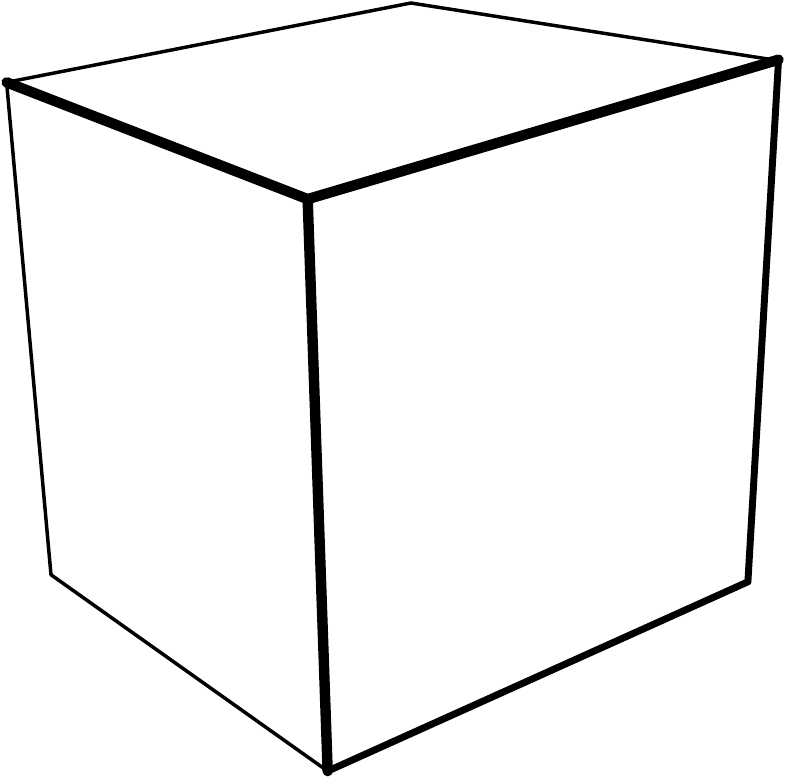}\end{minipage} & \begin{minipage}{0.6in}\includegraphics[width=\textwidth]{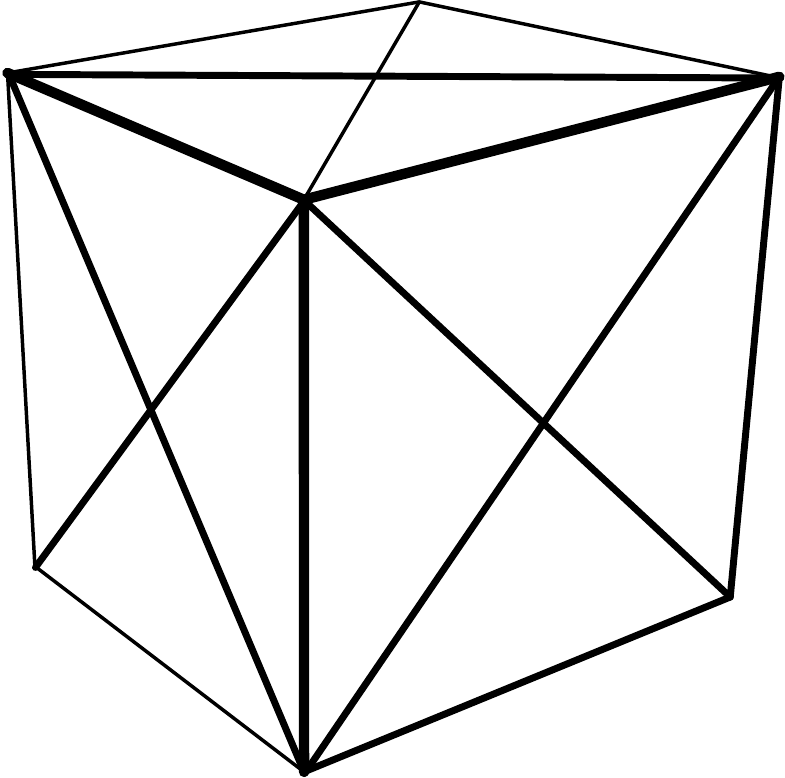}\end{minipage} & \begin{minipage}{0.57in}\includegraphics[width=\textwidth]{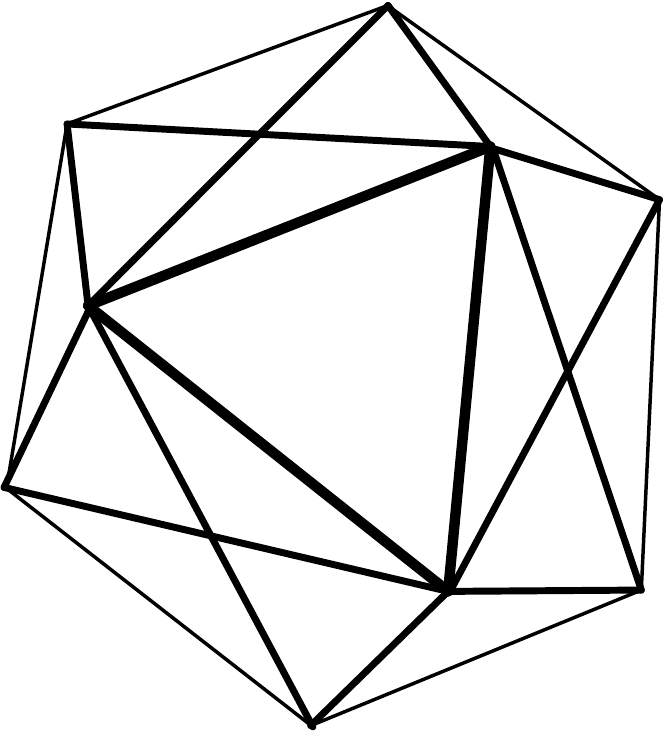}\end{minipage} & \begin{minipage}{0.6in}\includegraphics[width=\textwidth]{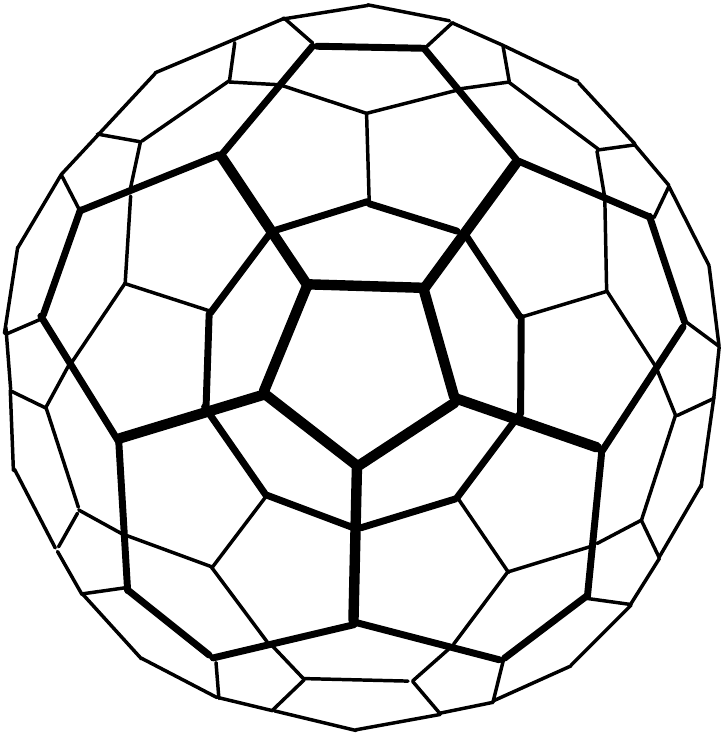}\end{minipage} & \begin{minipage}{0.6in}\includegraphics[width=\textwidth]{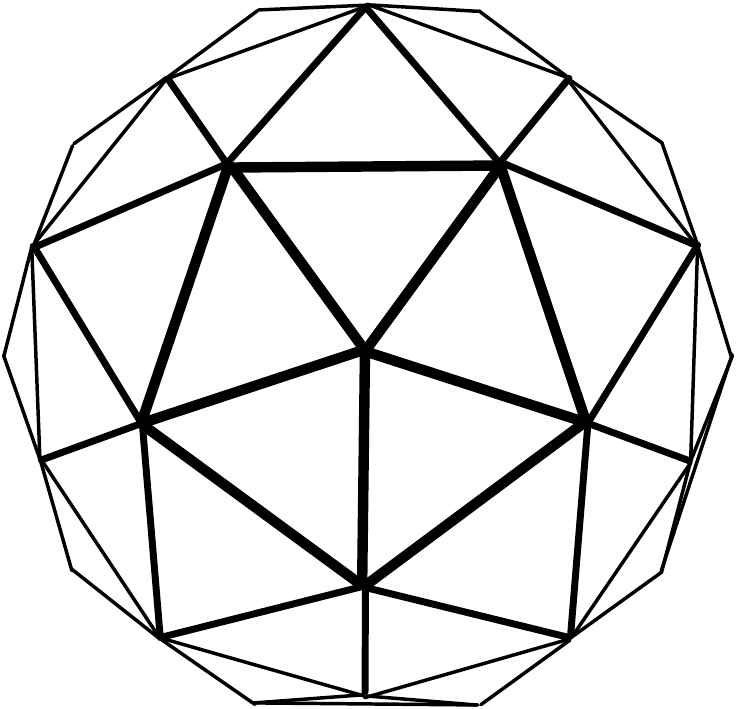}\end{minipage} \\
 &\hphantom{hexdecahedroid}&\hphantom{hexdecahedroid}&\hphantom{hexdecahedroid}&\hphantom{hexdecahedroid}&\hphantom{hexdecahedroid}&\hphantom{hexdecahedroid}\\
 \hline
 pentacope&1.0416667&1.0206207&1.0206207&1.1572751&1.1972885&1.1973624\\
 $m=5$& 1.3975425&1.1907107&1.2741137&1.2180011&1.2042164&1.2057998\\
 \hline
 hexdecahedroid &&1.0000000&1.0000000&1.1338934&1.1730984&1.1731707\\
 $m=8$&&1.4142136&1.4142136&1.4142136&1.2286820&1.1869546\\
 \hline
 tesseract &&&1.0000000&1.1338934&1.1730984&1.1731707\\
 $m=16$&&&1.4142136&1.4142136&1.2286820&1.1869546\\
 \hline
 octaplex &&&&1.2857143&1.3301686&1.33025058\\
 $m=24$&&&&1.4142136&1.3350119&1.34038194\\
 \hline
 tetraplex &&&&&1.3761599&1.3762447\\
 $m=120$&&&&&1.3776804&1.3842396\\
 \hline
 dodecaplex &&&&&&1.3763296\\
 $m=600$&&&&&&1.3834497\\
\hline
\end{tabular}
}
\caption{Maximum quantum values divided by local bounds $Q/L$ for Bell inequalities constructed from unit vectors pointing towards the vertices of four-dimensional (4D) Platonic solids. The upper and the lower numbers correspond to relative orientations giving the maximal and the minimal local bounds, respectively. In the first column the number of vertices for the solids is also shown. Orthographic projections of the vertices of the six 4D Platonic solids are also shown below the names.\label{table:d4}}
\end{center}
\end{table*}

{\bf Cross polytope--simplex}. Let there be a cross polytope on Alice's side and a simplex on Bob' side. For $d=3$ the simplex is the tetrahedron that we can get by appropriately eliminating half of the vertices of a cube. Therefore, $L$ is one half that of we have got for the cube, namely $L=4+4/\sqrt{3}=6.3094011$ while $Q/L=1.2679492$ is the same. For $d=4$ and for $d=5$ the optimally oriented simplex can also be determined from the numerical data. From those forms it follows that $L=3+\sqrt{5}(1+\sqrt{2})=8.3983456$, $Q/L= 1.1907107$, and $4(\sqrt{0.1}+\sqrt{0.3}+2\sqrt{0.6})=9.6525746$, $Q/L=1.2431916$ for $d=4$ and $d=5$, respectively. From the $d=4$ form we could give an ansatz for $d=6$, but it turned out that it can further be improved by applying a further rotation of a quite small angle in just a single plane defined by two basis vectors. This way we get $L=11.639817$ and $Q/L= 1.2027681$. For higher dimensions we have no reliable results. For the cross polytope-simplex pair we could not find any consistent behaviour that could be generalized.

{\bf Hypercube--hypercube}. Let us consider $d$-cubes on both sides. In three dimensions the matrix of an optimal transformation is:
\begin{equation}
\frac{1}{4}
\begin{pmatrix}
1/2 & \sqrt{5}+1 & \sqrt{5}-1\\
\sqrt{5}-1 & -1/2 & \sqrt{5}+1\\
-\sqrt{5}-1 & \sqrt{5}-1 & 1/2
 \end{pmatrix}.
\label{eq:cucub3dtr}
\end{equation}
Then $L=16(\sqrt{5}+1)/3=17.259029$ and $Q/L=4/(\sqrt{5}+1)=1.2360680$. In four dimensions the optimal transformation is the same as in the case of two cross polytopes or the cross polytope $4$-cube pair. If we eliminate appropriately one half of the simplices from both cubes, the Bell matrix we get will correspond to the sum of eight CHSH Bell matrices divided by $\sqrt{2}$. Like before, although they are not independent, there are classical strategies and measurement settings optimal for all of them. Therefore, $Q/L=\sqrt{2}$, and for the full cubes $L=64/\sqrt{2}$. For $d=5$ the smallest value we have got numerically was $L=165.01046$  ($Q/L=1.2411334$), and we could not recognize pattern leading to some analytical form. For $d=6$ the transformation we have found optimal was the same as the one for the cross polytopes. However, the Bell inequality this time did not come out as a sum of CHSH ones which could all be optimized by the same strategy, therefore the $L=4256/6/\sqrt{2}=501.57441$ value is slightly bigger than $4096/6/\sqrt{2}$, as it would be if the behaviour were the same as for $d=4$. Now $Q/L=128\sqrt{2}/133=1.3610476$.

{\bf Hypercube--simplex}. For a hypercube on one side and a simplex on the other in $3$ dimensions the $L$ and $Q$ values are just one half of the corresponding ones for the cube-cube pair, while $Q/L$ is the same. For $d>3$ we have only got numerical results. Actually, we think that with more calculations they could slightly be improved. The best values we got for $d=4$, $d=5$ and $d=6$ are $L=15.6971864$ ($Q/L=1.2741137$), $L=31.068002$ ($Q/L=1.2359984$) and $L=60.254415$ ($Q/L=1.2391900$), respectively.

{\bf Simplex--simplex}. For two simplices in $d=3$ (two tetrahedrons) the values of $L$ and $Q$ are one quarter of the ones for two cubes. For $d=4$ we have been able to figure out some analytical pattern from the numerical data, and we have got $L=2\sqrt{5}=4.4721360$, and $Q/L=5\sqrt{5}/8=1.3975425$. For $d=5$ and $d=6$ we have only numerical results, which probably could be improved with further calculations. What we have got is $L=5.8332724 $ ($Q/L=1.2342986$) and $L=6.6836480$ ($Q/L=1.2218876$) for $d=5$ and $d=6$, respectively.

{\bf Other pairs in three and four dimensions}. In Tables~\ref{table:d3} and \ref{table:d4} we show the results for the ratios of the quantum and the local bounds, both the maximum and the minimum ones for all pairs of Platonic bodies in $d=3$ and $d=4$, respectively. The maximum values are calculated as the products of the longest strategy vectors for the objects, as shown in the previous chapter, while the minimum ones are numerical results we got using the Nelder-Mead method starting from many random relative orientations. In Table~\ref{table:d3} the results for the cube and for the tetrahedron agree, as they must. One may notice in Table~\ref{table:d4} that $Q/L=\sqrt{2}$ when one of the solids is the octaplex, while the other one is either the hexdecahedroid (cross polytope), the octaplex or the tesseract (4-cube). In all cases a number of CHSH inequalities are involved, like in other similar cases. Let us take the 24 vectors one can get by considering all permutations and all sign variations of the coordinates of vector $(1,1,0,0)/\sqrt{2}$. These vectors correspond to the vertices of an octaplex. 
If the other body is the cross polytope in standard orientation, and we eliminate appropriately one half of the vertices of both bodies, the Bell matrix is  
\begin{equation}
\resizebox{\columnwidth}{!}{
$\left(\begin{array}{rrrrrrrrrrrr}%
 1  &   1  &   1  &   1  &   1  &  1  &   0  &  0  &   0 &   0  &   0  &   0 \\
 1  &  -1  &   0  &   0  &   0  &  0  &   1  &  1  &   1 &   1  &   0  &   0 \\
 0  &   0  &   1  &  -1  &   0  &  0  &   1  & -1  &   0 &   0  &   1  &   1 \\
 0  &   0  &   0  &   0  &   1  & -1  &   0  &  0  &   1 &  -1  &   1  &  -1 \\
\end{array}
\right)
$}
\label{eq:octaplexcpt}
\end{equation}
divided by $\sqrt{2}$. This involves six non-independent CHSH matrices such that there exist local strategies and measurement settings maximizing all of them. The inequality is the same as $I_{4,12}$ in Ref.~\cite{NTV14}. For the octaplex 4-cube pair the behaviour is similar. For two octaplexes one of the bodies has to be rotated by $\pi/4$ both in the $1-2$ and the $3-4$ planes. For two pentacopes the number of settings is five for both parties, and the  vectors pointing towards the vertices of the pentacopes in the four dimensional space are optimal measurement vectors. However, it turns out that one can find measurement vectors within three-dimensional subspaces giving the same quantum value. This means that the question raised in Ref.~\cite{vp09} about the existence of correlation-type Bell inequality with five settings for each of the two parties whose maximum violation requires four-dimensional measurement vectors still remains open. 

\section{Diagonally modified Bell inequalities}\label{belldiag}
In this section, we consider Platonic Bell inequalities in which the main diagonal entries of the Bell matrix, formed by scalar product of Euclidean vectors, are modified by subtracting the same constant from each diagonal entry. This modification of the original Platonic Bell inequalities has been performed in section~\ref{Qvalspec}, which, according to (\ref{eq:modem}), looks as follows:
\begin{equation}
M_{ij}(\lambda)=M_{ij}-\lambda\delta_{ij}=\vec A_i\cdot \vec A_j -\lambda\delta_{ij}.
\label{Mprime}
\end{equation}
In section~\ref{Qvalspec} we have also shown that the maximum quantum value, i.e., the Tsirelson bound~\cite{cirel80} becomes $(m_A^2/d)-\lambda m_A$, and this can be obtained with the same measurement vectors as without the modification. Note that in this case $M$ is a square matrix of size $m_A\times m_A$. Therefore, the Bell inequality defined by the expression in Eq.~(\ref{Mprime}) can still be regarded as a Platonic Bell inequality. In this way, we find Platonic Bell inequalities that are more robust to noise than the celebrated Clauser-Horne-Shimony-Holt Bell inequality~\cite{CHSH}, i.e., the quantum ($Q$) per local bound ($L$) is greater than $\sqrt 2$. However, this construction requires higher dimensional Hilbert spaces to reach the Tsirelson bound.

This diagonal modification technique has been applied previously by Davie and Reeds~\cite{DR} to obtain the best lower bound for the Grothendieck constant, $K_G\ge 1.67696$~\cite{Grothendieck,finch}. The Grothendieck constant of order $d$~\cite{Krivine}, denoted by $K_G(d)$, corresponds to the largest ratio $Q(M,d)/L(M)$ among arbitrary $M$ real matrices, where $Q$ is defined by $Q(M,d)=\sum_{i,j}M_{i,j}\vec a_i\cdot\vec b_j$, where $\vec a_i,\vec b_j$ are unit vectors in the $d$-dimensional Euclidean space. This relation was first noticed by Tsirelson~\cite{Tsirelson87}.
Note that $K_G=\lim_{d\rightarrow\infty}K_G(d)$. Therefore the lower bound $K_G\ge 1.67696$ corresponds to a correlation-type Bell expression $M$ with $Q(M,d)/L(M)\ge 1.67696$. However, the constructed Bell inequality has an infinite number of settings on both Alice's and Bob's side. We note that the exact value of $K_G(d)$ is known only for $d=2$, $K_G(2)=\sqrt 2$ with the Bell matrix corresponding to the CHSH inequality. This proof is due to Krivine~\cite{Krivine}.

The first example of $M$ with modest size exceeding the ratio $Q/L=\sqrt 2$ was discovered by Reeds and Sloane in 1990. They used an $M$ matrix of size $120\times 120$ of the family~(\ref{Mprime}) with $\lambda=0$, where $\vec A_i$ are formed by the 240 eight-dimensional unit vectors of the lattice $E_8$ resulting in $K_G(8) \ge 45/31 = 1.4516129$. A finite $M(\lambda)$ matrix size in (\ref{Mprime}), where $\lambda>0$, was first used by Fishburn and Reeds~\cite{fishburn}. Their $20\times 20$ matrix $M(\lambda)$ provides a ratio $Q/L=10/7$, which implies that $K_G(5) \ge 10/7 = 1.4285714$. Note, however, that the construction by Fishburn and Reeds \cite{fishburn} does not give a lower bound on $K_G(4)$ greater than $\sqrt 2$. 

In the following, we prove that 
\begin{equation}
K_G(4) \ge \frac{70}{5+27\phi}-4\times 10^{-6}= 1.4377539
\label{KG4LB}
\end{equation}
using Platonic Bell inequalities, where $\phi$ is the golden ratio and the dimension of the matrix $M(\lambda)$ is $60\times 60$. Note that there exist better lower bounds in the literature. Namely, $K_G(4)\ge 1.445207$ ~\cite{vertesi}, $K_G(4)\ge 1.44566$~\cite{hua} and $K_G(4)\ge 1.4821664$~\cite{peter17}. However, neither of them is based on Platonic Bell inequalities. Let us also mention the upper bound $K_G(4)\le \pi/2$~\cite{Krivine}.

To get the ratio $Q/L = 1.4377539$ using the Platonic Bell inequalities, we construct $M(\lambda)$ according to Eq.~(\ref{Mprime}). In particular, we choose $\{\vec A_i\}$ as the vertices of the halved tetraplex ($m_A=60$, $d=4$). We explicitly define the 60 vertices $\vec A_i$ as follows:
Four vertices come from
\begin{equation}
(0,0,0,1)
\end{equation}
by permuting the coordinates, eight vertices from 
\begin{equation}
(1/2,\pm 1/2,\pm 1/2,\pm 1/2)
\end{equation}
and the remaining 48 vertices are obtained by taking all even permutations of 
\begin{equation}
(\phi/2,\pm 1/2,\pm 1/(2\phi),0),
\end{equation}
where $\phi$ is the golden ratio. Note that the 120-vertex regular tetraplex comprises the above $\vec A_i$ vertices and its antipodal ones $-\vec A_i$. 
The quantum value $Q(\lambda)$ for the Bell matrix $M(\lambda)$ in (\ref{Mprime}) is 
\begin{equation}
Q(\lambda)=m_A^2/d-\lambda m_A = 450-60\lambda
\label{Qlambda}
\end{equation}
if $\lambda\le m_A/(2d)=15/2$. This is due to the analysis of Section~\ref{Qvalspec}, in which case the Tsirelson bound of the Bell inequality is given by the 60 vertices of the halved tetraplex shown above. Indeed, the 120 vertices of the tetraplex readily obeys the constraints of the Observation in Section~\ref{Qvalspec}. Moreover, the set of 120 vertices is centrally symmetric (i.e. every $\vec A_i$ has an antipodal vector $-\vec A_i$ in the set). Therefore, by the argument at the end of Section~\ref{Qvalspec} the halved 60-vertex tetraplex also obeys the constraints of the Observation and the maximum quantum bound is given by (\ref{Qlambda}).

\begin{figure}[t!]
\begin{center}
\includegraphics[trim=5 10 0 -10,clip,width=8.1cm]{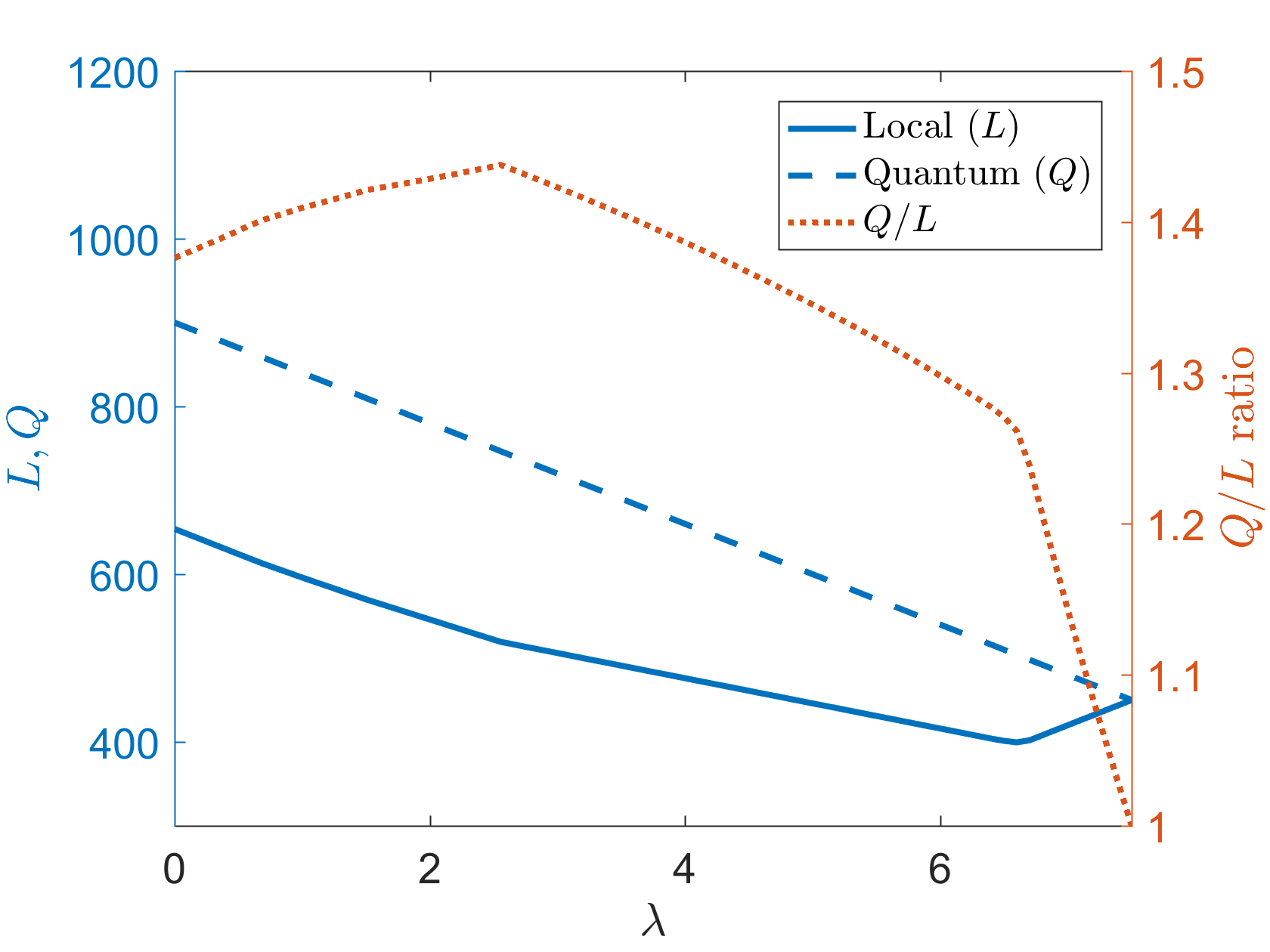} 
\end{center}
\caption{The $Q/L$ ratio (red dots) in the function of $\lambda$ for the Platonic Bell inequality defined by the coefficients~(\ref{Mprime}), where $\vec A_i$ are the vertices of the halved tetraplex. The functions $Q$ (blue dashed line) and $L$ (blue solid line) are also plotted against $\lambda$. The maximum ratio $Q/L=1.4377579-\epsilon$ is obtained at $\lambda=23/9$, where $\epsilon=4\times 10^{-6}$. The value $1.4377579$ is analytically given by $70/(5+27\phi)$, where $\phi=(1+\sqrt 5)/2$ is the golden ratio.} \label{fig_golden}
\end{figure}

However, the local bound is difficult to calculate. Recall that, for a generic $M$ matrix the calculation of the local bound~(\ref{eq:localval}) is an instance of $K_{m,n}$-quadratic programming~\cite{raga}. In our case the matrix $M$ is replaced by the one-parameter family of matrices $M(\lambda)$ in (\ref{Mprime}). However, unlike the case $\lambda=0$ discussed in Section~\ref{Lbound}, we have not found an efficient method to compute the local bound for $\lambda>0$. On the other hand, a brute-force computation based on the enumeration of local deterministic strategies would require the evaluation of $2^{m_A-1}=2^{59}$ instances. This task is impossible to perform in a reasonable time even with supercomputers. However, the algorithm based on the branch-and-bound (BB) technique~\cite{BB} developed in Ref.~\cite{peter17} makes the problem tractable for our particular case. This code is publicly available~\cite{github}, and supports parallel computation of the local bound~(\ref{eq:localval}). With the latest parallel implementation available in ~\cite{github}, running our problem on a 56-core workstation, the local bound $L(\lambda)$ is obtained for an arbitrary $\lambda$ in about 2 minutes. The obtained ratio $Q/L=(450-60\lambda)/L(\lambda)$ as a function of $\lambda$ is plotted from $\lambda=0$ to $\lambda=7.5$ in Fig.~\ref{fig_golden}. For comparison, we also plot $Q(\lambda)$ and $L(\lambda)$ in the same figure. The abrupt change in the $L(\lambda)$ function is observed just as the ratio $Q(\lambda)/L(\lambda)$ reaches its peak value at $\lambda=23/9$. 

The BB-based code~\cite{github} computes exactly the local bounds of Bell matrices whose coefficients are all integers. 

Some technical notes on the computation of the local bound $L(\lambda)$ are in order. The BB-based code~\cite{github} computes with exact arithmetics the local bound of Bell matrices whose entries are all integers. Hence, for a matrix $M$ with integer entries the calculated local bound is exact. However, the entries of our special $M(\lambda)$ matrices are in general not integers. We avoid this problem by the following method. This way we obtain a very good upper bound on $L(M(\lambda))$ as follows. We multiply $M(\lambda)$ by a large number ($10^6$, in our particular case) and round each entry to the nearest integer below it. That is, we have $10^6M=M_{\rm int}+\Delta M$, where $M_{\rm int}=\lfloor 10^6M\rfloor$ is an integer matrix, and the entries of $\Delta M$ are $\Delta M_{ij}\in [0,1]$. The number $10^6$ has been chosen in order not to occur overflow, as the code is limited to Bell matrices where $\sum_{i,j}|M_{ij}|\le 2^{31}$. Then we have the following upper bound
\begin{equation}
L(10^6M)\le L(M_{\rm int})+L(\Delta M),
\label{10o6M}
\end{equation}
where we used that for any two Bell matrices we have $L(M_1+M_2)\le L(M_1) + L(M_2)$. Furthermore, due to all the positive entries in $\Delta M_{i,j}$, we have $L(\Delta M)=\sum_{i,j}|\Delta M_{ij}|$. Then, if we divide both sides of (\ref{10o6M}) by $10^6$, we get the following upper bound on $L(M)$:
\begin{equation}
L(M)\le \frac{L(M_{\rm int})}{10^6}+\frac{\sum_{i,j}|\Delta M_{ij}|}{10^6},
\end{equation}
where $L(M_{\rm int})$ is the output of the BB-based code~\cite{github}. The maximum $Q/L$ ratio of $M(\lambda)$ occurs at $\lambda=23/9$ (see Fig.~\ref{fig_golden}). In this case, we have $L(M_{\rm int})=519326904$ and $\sum_{i,j}|\Delta M_{ij}|=1622.88706865$, which results in an upper bound $L(M)\le 519.32852689$ and a lower bound $Q(M)/L(M)\ge 1.4377539$, where the latter value equals $70/(5+27\phi)-\epsilon$, where $\epsilon = 4\times 10^{-6}$. In summary, using measurement directions pointing to the vertices of the four-dimensional 60-vertex halved tetraplex and the diagonal modification of the Bell matrix $M$, we obtain the ratio $Q/L\ge 1.4377539$ and the corresponding lower bound $K_G(4)\ge 1.4377539$.

\section{Discussion}\label{disc}
Tavakoli and Gisin~\cite{TavGis} constructed Bell inequalities with coefficients $M_{ij}=\vec A_i\cdot\vec B_j$, where $\vec A_i$ and $\vec B^*_j$ are the unit vectors pointing to the desired measurement directions on the sides of Alice and Bob, respectively, i.e., towards the vertices of Platonic bodies. Then the obtained quantum value of the Bell expression with these measurement directions is $\sum_{ij}(\vec A_i\cdot\vec B_j)^2$. Using the NPA method~\cite{npa}, the authors find that the above measurement directions provide optimal quantum violation (i.e. the Tsirelson bound~\cite{cirel80}). Such constructions have been called Platonic Bell inequalities. Motivated by this construnction, we generalized the Platonic Bell inequalities to all possible dimensions. Due to a theorem of Tsirelson~\cite{Tsirelson87}, the generalization of the construction of the Bell inequalities to higher dimensional spaces does make sense, the result will correspond to a genuine Bell scenario, however, the symmetry of the solids will show up in the abstract Tsirelson space. To prove that this generalization is possible, we have shown that all Platonic solids (and Archimedean solids) have some property that ensures this, and that the form of the quantum bound is very simple and independent of the relative orientation of the two solids on Alice and Bob side. However, the local bound, and hence the maximum violation of the inequality depends on this, which is discussed in detail in this paper. 

In more than two dimensions some of the Platonic solids have too many vertices, hence computing the local bounds of the Bell inequalities constructed from them in a brute force way may not be feasible. We have found an efficient method which makes it possible to calculate the local bound exactly in all cases where the Bell coefficients are given as scalar products of low-dimensional Euclidean vectors, even when the number of measurement settings is large. The practical significance of the method is demonstrated by the exact computation of the local bound for the 300-setting four-dimensional Platonic Bell inequality. The Euclidean vectors in this method do not have to be unit vectors, which allows us to prove that the method can be used not only for Platonic Bell inequalities but also for all bipartite two-outcome Bell inequalities where the rank of the Bell matrix is low. 

Finally, we study Platonic Bell inequalities where the main diagonal entries of the Bell matrix formed by the scalar product of Euclidean vectors are modified by a constant term. In this way, we find Platonic Bell inequalities that are more robust to noise than the celebrated Clauser-Horne-Shimony-Holt Bell inequality, that is, the quantum ($Q$) per local bound ($L$) is greater than $\sqrt 2$. However, this construction requires higher dimensional Hilbert spaces to achieve maximal quantum violation. In this regard, we leave open the question whether the Tsirelson bound with a $Q/L$ ratio larger than $\sqrt 2$ can be achieved by a suitable symmetric Bell inequality using a two-qubit maximally entangled state where the measurements point to the vertices of some three-dimensional symmetric polyhedron.

\emph{Acknowledgements.} We thank Péter Diviánszky for his help in writing the parallel version of the $K_{m,n}$ quadratic programming code and Marco T\'ulio Quintino for valuable discussions. TV acknowledges the support of the EU (\mbox{QuantERA} eDICT) and the National Research, Development and Innovation Office NKFIH (No. 2019-2.1.7-ERA-NET-2020-00003).


\appendix

\section{Relationship between the Euclidean and the Hilbert space description}\label{EuHil}

In the present paper the quantum measurements performed by the parties are represented by unit vectors of an Euclidean space. However, quantum states live in Hilbert space and measurements correspond to appropriate Hermitian operators in that Hilbert space. As shown for example in Ref.~\cite{TavGis} and discussed briefly in the main text, if the quantum system is a pair of qubits in the maximally entangled $|\Phi^+\rangle=(|00\rangle+|11\rangle)/\sqrt{2}$ state, the quantum measurements performed by the parties on their respective qubits can be represented by vectors in a three-dimensional Euclidean space such that the quantum expectation values occuring in the Bell-expression are simply the scalar products of the corresponding Euclidean vectors. Moreover, if the qubits are spins, those vectors are very simply related to orientations of the devices measuring the spin projection in the physical space, namely for one of the parties they point towards the measurement directions themselves, while for the other party one can get the measurement directions by reflecting the vectors through the $x-z$ plane. We note that a similar relationship exists if the state is the singlet one $|\Psi^-\rangle=(|10\rangle-|01\rangle)/\sqrt{2}$. Then the measurement vectors and the measurement directions point to the opposite directions for both parties. In Ref.~\cite{bolonek} this latter relationship has been exploited. A beautiful feature of the Platonic Bell inequalities in three dimensions is that if the system is a pair of one-half spins, the symmetries of the Platonic solids show up directly in the orientations of the measurement settings in the physical space. Unfortunately, this beauty is lost if the system is not like that (even if it is a pair of qubits, but not spins). Therefore for the higher dimensional generalization discussed in the present paper the symmetries are only present in an abstract space, not in the physical arrangements concerning the measurement settings.

However, it is not even obvious that the considerations in more than three dimensional spaces are relevant at all for Bell scenarios. A theorem by Tsirelson~\cite{Tsirelson87} ensures that this is the case. Due to this theorem, for any Euclidean space there exists a Hilbert space $H$ and a state $|\Phi\rangle\in H\otimes H$ such that for any unit vectors $\vec a$ and $\vec b$ in the Euclidean space there exist binary measurement operators $\hat a$ and $\hat b$ in the Hilbert space such that
\begin{equation}
\vec a\cdot\vec b=\langle\Phi|\hat a\otimes\hat b|\Phi\rangle.
\label{eq:vecop}
\end{equation}
Hence, for each party, any unit vector does represent some quantum measurement performed on the respective subsystem of the quantum system $|\Phi\rangle$. The statement can be proven by construction \cite{acin06,vp08}. The considerations we will give here are based on this construction.

Let us consider a $d$-dimensional Eulidean space. If there exist $d$ mutually anticommuting operators $\hat\gamma_i$ ($i=1,\dots,d$) whose square is the identity operator (that is $\hat\gamma_i\hat\gamma_j+\hat\gamma_j\hat\gamma_i=2\delta_{ij}\hat I$) in the $D$-dimensional Hilbert space, then with $\hat a=\sum_{i=1}^d a_i\hat\gamma_i$, $\hat b=\sum_{i=1}^d b_i\hat\gamma_i^t$ ($t$ denotes the transposition) and with the maximally entangled state $|\Phi\rangle=\sum_{\mu=1}^D|\mu\mu\rangle/\sqrt{D}$ Eq.~(\ref{eq:vecop}) is satisfied. Indeed, 
$\langle\Phi|\hat a\otimes\hat b|\Phi\rangle={\rm Tr}(\hat a\hat b^t)/D=\sum_{i,j=1}^d a_ib_j{\rm Tr}(\hat\gamma_i\hat\gamma_j)/D=\vec a\cdot\vec b$. 
The last equality follows from the fact that $\hat\gamma_i$ and $\hat\gamma_i$ anticommute if $i\neq j$, therefore ${\rm Tr}(\hat\gamma_i\hat\gamma_j)=0$, and that $\hat\gamma_i^2$ is the identity operator whose trace is $D$. It can similarly be shown that the square of both $\hat a$ and $\hat b$ is the identity operator, therefore they do represent binary measurements with outcomes $+1$ and $-1$, as required.

One can construct a set of $2n+1$ $\hat\gamma_i$ operators as tensor products of $n$ Pauli operators and two-dimensional identity operators $\hat I_2$ in appropriate orders. An example of such a construction for arbitrary $n$ is given explicitly in the Appendix of \cite{vp08}. This way one can get the $d$ $\hat\gamma_i$ operators needed to assign quantum measurements to the unit vectors of the $d$-dimensional Euclidean space as tensor products of $\lfloor d/2\rfloor$ operators acting on qubit spaces. Therefore, the Hilbert space $H$ of each subsystem has $D=2^{\lfloor d/2\rfloor}$ dimensions.

For $d=3$ the assignment is the same as we have already discussed. The three $\hat\gamma_i$ operators are the Pauli operators themselves, and $D=2$. For $d=2$ one can choose any two of the Pauli operators. With the choice of $\hat\sigma_x$ and $\hat\sigma_z$, one arrives at a real Hilbert space. For $d=5$ the five operators corresponding to the recipe given in Ref.~\cite{vp08} are $\gamma_1=\hat\sigma_x\otimes\hat I_2$, $\gamma_2=\hat\sigma_y\otimes\hat I_2$, $\gamma_3=\hat\sigma_z\otimes\hat\sigma_x$, $\gamma_4=\hat\sigma_z\otimes\hat\sigma_y$ and $\gamma_5=\hat\sigma_z\otimes\hat\sigma_z$, and $D=4$. For $d=4$ any four of these operators may be chosen. However only three of them are real, therefore we can not get a real Hilbert space this way. Even if we do not follow the recipe, we can not find four pairwise anticommuting operators among all real ones which are tensor products of either two Pauli operators or one Pauli operator and the identity.

Following the construction discussed above, the dimensionality of the Hilbert space grows exponentially with $d$. We do not know the conditions for the existence of some more economical assignment. In Ref.~\cite{vp09} almost a hundred Bell inequalities that required four dimensional Euclidean vectors for maximum violation have been considered, and neither of them could maximally be violated in smaller than $4\times 4$ dimensional complex Hilbert spaces. We have also investigated numerically some inequalities requiring $6$ and $8$ dimensional Euclidean spaces for maximum violation, and we could not violate them maximally with less than $8\times 8$ and $16\times 16$ dimensional quantum systems, respectively.

It has been shown that for all $d$-dimensional vector pairs $\vec a$ and $\vec b$ one can assign operator pairs $\hat a$ and $\hat b$ from the $D=2^{\lfloor d/2\rfloor}$ dimensional Hilbert space $H$ such that Eq.~(\ref{eq:vecop}) is satisfied with $|\Phi\rangle=\sum_{\mu=1}^D|\mu\mu\rangle$. The opposite statement is not true for $d>3$. As all measurement operators constructed have zero trace, other operators belonging to legitimate two-outcome measurements are excluded. Not even all zero trace operators are covered: the $d-1$ independent parameters characterizing the unit vectors are not sufficient to characterize all such operators. Moreover only Hilbert spaces with dimensionalities of a power of two have occured.

Therefore, to show that for any quantum scenario one can find appropriate Euclidean vectors for the two parties requires a different construction. This relation is less important as far as this paper is concerned, but for the sake of completenes we still sketch the argument here. Let $H_A$ and $H_B$ be the $D_A$ and $D_B$ dimensional Hilbert spaces of the two parties, respectively. Let the measurements be performed on the state $|\Phi\rangle\in H_A\otimes H_B$. Then for any $\hat a$ and $\hat b$ for the parties the $2D_AD_B$ dimensional real vectors $\vec a$ and $\vec b$ whose components are the real and the imaginary parts of the components of $\hat a\otimes\hat I_B|\Phi\rangle$ and $\hat I_A\otimes\hat b|\Phi\rangle$, respectively satisfy Eq.~\ref{eq:vecop}, where $I_A$ and $I_B$ are the identity operators in $H_A$ and $H_B$, respectively. Using $\hat a^2=\hat I_A$ and $\hat b^2=\hat I_b$, the statement is easy to prove. If the number of settings is $m_A$ and $m_B$ for the respective parties, then an $(m_A+m_B)$-dimensional subspace is sufficient, as no more vectors are involved. Moreover, if one is only interested in the maximum violation, no more than an $m_A$ or $m_B$ dimensional subspace is needed, whichever is the smaller. The reason is that the optimum vectors on one side are in the subspace spanned by the vectors on the other side. 

\section{Proof of special property}\label{specpropproof}

In Section~\ref{Qvalspec} we have proven that the maximum quantum value for the Bell inequality whose Bell coefficients have been constructed as the scalar products of two sets of $d$-dimensional unit vectors $\vec A_i$ ($i=1,\cdots,m_A)$ and $\vec B_j$ ($j=1,\cdots,m_B)$ is $m_Am_B/d$ if both sets satisfy the constraints give in the Observation. This means that the columns of matrix $A_{ij}$ whose rows are the coordinates of vectors $\vec A_i$ are orthogonal to each other and have equal norm, which is $\sqrt{m_A/d}$, and $\vec B_j$ also has this property. Here we will prove that cross polytopes, simplices and $d$-cubes satisfy the constraints for all $d$, and it is also true for the 2-dimensional regular polygones. The additional two three-dimensional and the three  four-dimensional Platonic bodies has simply been checked, we do not consider them in this Appendix.

The cross polytope is simple. We get its matrix $A_{ij}$ if we write the $d$-dimensional unit matrix above minus one times the same matrix. Then the columns are trivially orthogonal and their norm is $\sqrt{2}$ (There is one $1$ and one $-1$ in each column), as it should be.

Let us consider simplices in $d$-dimensions. They have $d+1$ vertices. Let the elements of the matrix of size $(d+1)\times d$ be $A_{ij}=0$ for $j>i$, $A_{ij}=\tau_j$ for $j=i$ and  $A_{ij}=-\sigma_i$ for $j<i$, that is:
\begin{equation}
\hat A = 
\begin{pmatrix}
\tau_1 & 0 & 0 & 0 & 0 & \cdots \\
-\sigma_1 & \tau_2 & 0 & 0 & 0 & \cdots \\
-\sigma_1 & -\sigma_2 & \tau_3 & 0 & 0 & \cdots \\
-\sigma_1 &-\sigma_2 & -\sigma_3 & \tau_4 & 0 & \cdots \\
-\sigma_1 &-\sigma_2 &-\sigma_3 & -\sigma_4 & \tau_5 & \cdots \\
-\sigma_1 &-\sigma_2 &-\sigma_3 & -\sigma_4 & -\sigma_5 & \cdots\\
\vdots & \vdots & \vdots & \vdots & \vdots & 
 \end{pmatrix}
\label{eq:simpexplic}
\end{equation}
where
\begin{align}
\tau_j&=\sqrt{\frac{(d+1)(d+1-j)}{d(d+2-j)}}\nonumber\\
\sigma_j&=\sqrt{\frac{d+1}{d(d+1-j)(d+2-j)}}.
\label{eq:tausigma}
\end{align}
it is easy to verify that $\tau_1=1$, and that $\tau_j^2=\tau_{j+1}^2+\sigma_j^2$. From these it follows that
\begin{equation}
1=\tau_1^2=\sigma_1^2+\tau_2^2=\sigma_1^2+\sigma_2^2+\tau_3^2=\cdots=\sum_{j=1}^{i-1}\sigma_j^2+\tau_i^2,
\label{eq:simplexnorm}
\end{equation}
which is nothing else than the squares of the norms of the rows. It follows from Eq.~(\ref{eq:simpexplic}) that the scalar product of the $i$th and the $k$th rows for any $k>i$ can be written as
\begin{equation}
\sum_{j=1}^{i-1}\sigma_j^2-\sigma_i\tau_i=1-(\tau_i^2+\sigma_i\tau_i)=1-\frac{d+1}{d}=-\frac{1}{d},
\label{eq:scalprodrows}
\end{equation}
where we used Eq.~(\ref{eq:simplexnorm}) and that $\tau_i^2+\sigma_i\tau_i=(d+1)/d$, which can be verified using Eq.~(\ref{eq:tausigma}). From these it follows that the $d+1$ rows of the matrix are the coordinates of normalized vectors pointing towards the vertices of a $d$-dimensional simplex. Now, let us look at the columns to show that the constraints are met. The square of the norm of the $j$th column is $\tau_j^2+(d+1-j)\sigma_j^2=(d+1)/d$, as required, which can be checked by using Eq.~(\ref{eq:tausigma}). It can also be checked that $\tau_j+(d+1-j)\sigma_j=0$, that is the sum of the elements in each column is zero. From Eq.~(\ref{eq:simpexplic}) it is obvious that the scalar product of columns $j$ and $l$ where $(l>j)$ is the sum of the elements of column $l$ multiplied by $\sigma_j$, therefore, the columns are orthogonal to each other, so all constraint are obeyed.

A simple way to generate the components of the $2^d$ vectors pointing towards the simplices of a $d$-cube is to take the binary representation of the integers between zero and $2^d-1$ (writing the necessary number of zeros in front of the binary numbers), and replacing every zero by $-1$. If one wants the vectors normalized, they should be multiplied by $1/\sqrt{d}$. To show that the columns of the matrix whose rows are these vector components have equal norm and they are orthogonal to each other, we may forget about the normalization. Then the last column will have $-1$ and $+1$ values alternating, starting with a $-1$. The column before that will have pairs of $-1$ and $+1$ values alternating. In each column periods of $-1$s and $+1$s of equal length alternate, where the length is some power of two, it is $2^{d-1}$ for the first column (the upper half of the column is -1, the lower one is +1) and $2^0=1$ for the last one. It is easy to see that the scalar product of any two different columns is zero: in each interval corresponding to a section of equal signs in the column that has the longer sections there is an even number of sections in the other column, therefore there are the same number of $+1$ and $-1$ values in every such interval. The equality of the norms of the column is trivial, all elements have the same absolute value.

Let us look at the Platonic solids of two-dimensional spaces, the convex regular polygons. Let $\vec A_i$ be unit vectors pointing towards the vertices of the regular $m$-sided polygon, and $\vec x$ and $\vec y$ be two orthogonal vectors. Then
\begin{align}
\sum_{i=1}^m (\vec A_i\cdot \vec x)(\vec A_i\cdot \vec x)&=\sum_{i=1}^m\cos^2(\varphi_i)\nonumber\\
&=\sum_{i=1}^m\frac{\cos(2\varphi_i)+1}{2}=\frac{m}{2}\nonumber\\
\sum_{i=1}^m (\vec A_i\cdot \vec x)(\vec A_i\cdot \vec y)&=\sum_{i=1}^m\cos(\varphi_i)\sin(\varphi_i)\nonumber\\
&=\sum_{i=1}^m\frac{\sin(2\varphi_i)}{2}=0,
\label{eq:twodcontr}
\end{align}
where $\varphi_i$ the angle between $\vec A_i$ and $\vec x$ (measured towards $\vec y$). A sufficient condition for the equations above to be satisfied is that vectors $\vec A'_i$, whose angles with $\vec x$ are two times the ones between $\vec A_i$ and $\vec x$ have a $k$-fold symmetry. This is true in the present case, $k=m$ and $k=m/2$ for odd and even sided polygons, respectively. There are cases when $\vec A$ has no such symmetry at all, while $A'_i$ does (for example, when $m$=2 and $\vec A_1$ and $\vec A_2$ are two orthogonal vectors). We get the constraints of the Observation if we write the basis vectors into the equations.

The considerations above have relevance if $d>2$ as well, this is the reason we have written the equations in the more general form. Let  $\vec A_i$ have an $l$-fold ($l>2$) symmetry for rotations in some plane. Let $\vec x$ and $\vec y$ be orthogonal vectors in that plane. Then Eq.~(\ref{eq:twodcontr}) is modified as 
\begin{align}
\sum_{i=1}^m (\vec A_i\cdot \vec x)(\vec A_i\cdot \vec x)&=\sum_{i=1}^m\pi_i\cos^2(\varphi_i)\nonumber\\
&=\sum_{i=1}^m\frac{\pi_i(\cos(2\varphi_i)+1)}{2}=\sum_{i=1}^m\frac{\pi_i}{2}\nonumber\\
\sum_{i=1}^m (\vec A_i\cdot \vec x)(\vec A_i\cdot \vec y)&=\sum_{i=1}^m\pi_i\cos(\varphi_i)\sin(\varphi_i)\nonumber\\
&=\sum_{i=1}^m\frac{\pi_i\sin(2\varphi_i)}{2}=0,
\label{eq:twodcontrhigd}
\end{align}
where $\pi_i$ is the length of the projection of $\vec A_i$ onto the plane and $\varphi_i$ is the angle between the projection and $\vec x$. Therefore, such symmetry ensures that some constraints are satisfied.	

\end{document}